\documentclass{JHEP3} % 10pt is ignored!
\newcommand\ee{\end{equation}}

\JHEPspecialurl{http://jhep.sissa.it/JOURNAL/JHEP3.tar.gz}

\usepackage{amssymb,epsfig,multicol,multicol,bbm}

\newcommand\fverb{\setbox\fverbbox=\hbox\bgroup\verb}
\newcommand\fverbdo{\egroup\medskip\noindent%
                        \fbox{\unhbox\fverbbox}\ }
\newcommand\fverbit{\egroup\item[\fbox{\unhbox\fverbbox}]}
\newbox\fverbbox

\usepackage{amsmath}

\usepackage{epsfig}

\def\p{\partial}                                % partial derivative
\def\slashchar#1{\setbox0=\hbox{$#1$}                   % set a box for #1
   \dimen0=\wd0                                         % and get its size
   \setbox1=\hbox{/} \dimen1=\wd1                       % get size of /
   \ifdim\dimen0>\dimen1                                % #1 is bigger
      \rlap{\hbox to \dimen0{\hfil/\hfil}}              % so center / in box
      #1                                                % and print #1
   \else                                                % / is bigger
      \rlap{\hbox to \dimen1{\hfil$#1$\hfil}}           % so center #1
      /                                                 % and print /
   \fi}

\newcommand{\beq}{\begin{equation}}
\newcommand{\eeq}{\end{equation}}
\newcommand\be{\begin{equation} }
\newcommand\bea{\begin{eqnarray}}
\newcommand\eea{\end{eqnarray}}

\def\endtitle{\par\end{quotation}\vskip3.5in minus2.3in\newpage}

     \def\x{\hat x}

       \def\P{\Pi}

 \title{
 \begin{center}
 \bf  \Large  Scalar field propagation in the $\phi^4$ $\kappa$-Minkowski model
 \end{center}
 }

\author{S. Meljanac$^1$, A. Samsarov$^1$, J. Trampeti\'c$^{1,2}$ and M. Wohlgenannt$^3$ \\ 
1. Rudjer Bo\v skovi\' c Institute, 
P.O.Box 180, HR-10002 Zagreb, Croatia \\
2. Max-Planck-Institut f\"ur Physik, (Werner-Heisenberg-Institut),
  	 F\"ohringer Ring 6, D-80805 M\"unchen, Germany\\
3. Faculty of Physics, University of Vienna, Boltzmanngasse 5, A-1090 Vienna, Austria\\
E-mail: \email{meljanac@irb.hr}, \email{asamsarov@irb.hr}, \email{josipt@rex.irb.hr }
\email{michael.wohlgenannt@univie.ac.at}}

\abstract{ 
In this article we use the noncommutative (NC) $\kappa$-Minkowski $\phi^4$ model
based on the $\kappa$-deformed star product, (${\star}_h$). 
The action is modified by expanding up to linear order in 
the $\kappa$-deformation parameter $a$, 
producing an effective model on commutative spacetime.  
For the computation of the tadpole diagram contributions 
to the scalar field propagation/self-energy, we anticipate
that statistics on the $\kappa$-Minkowski is specifically $\kappa$-deformed.  
Thus our prescription in fact represents {\it hybrid} approach
between standard quantum field theory (QFT) and NCQFT on the
$\kappa$-deformed Minkowski spacetime, resulting in a $\kappa$-effective model.
The propagation is analyzed in the framework of the two-point Green's function
for low, intermediate, and for the Planckian propagation energies, respectively.
Semiclassical/hybrid behavior of the first order quantum correction do show up
due to the $\kappa$-deformed momentum conservation law. 
For low energies, the dependence of the tadpole contribution on the deformation 
parameter $a$ drops out completely, while
for Planckian energies, it tends to 
a fixed finite value. The mass term of the scalar field is shifted and 
these shifts are very different at different propagation energies. 
At the Planckian energies we obtain the direction dependent 
$\kappa$-modified dispersion relations. 
Thus our $\kappa$-effective model for 
the massive scalar field shows a birefringence effect.
}


\begin{document}
%\maketitle
 
%\newpage

%%%%%%%%%%%%%%%%%%%%%%%%%%%%%%%%%%%%%%%%%%%%%%%%%%%%%%%%%%%%%%%%%%%
%%%%%%%%%%%%%%%%  Introduction %%%%%%%%%%%%%%%%%%%%%%%%%%%%%%%%%%%%%this model
%%%%%%%%%%%%%%%%%%%%%%%%%%%%%%%%%%%%%%%%%%%%%%%%%%%%%%%%%%%%%%%%%%%

\section{Introduction}

Recently, it was established that if the 
$\kappa$-Poincar\'{e} Hopf algebra is assumed to describe 
physics on $\kappa$-Minkowski spacetime 
\cite{Lukierski:1991pn,Lukierski:1992dt,Majid:1994cy}, then it is
necessary to accept certain modifications in statistics obeyed by the
particles. This means that $\kappa$-Minkowski spacetime leads to a
modification of particle statistics which results in deformed oscillator algebras
\cite{Govindarajan:2009wt,Young:2007ag,Arzano:2007ef,Daszkiewicz:2007ru}.
A deformation of the Poincar\'{e} algebra can be 
performed by means of the twist operator
\cite{drinfeld,drinfeldN,Borowiec:2004xj,Borowiec:2006fc,Balachandran:2007vx} 
which happens to include the dilatation generator. And thus belongs to the universal
enveloping algebra of the general linear algebra
\cite{Bu:2006dm,Govindarajan:2008qa,Borowiec:2008uj,Bu:2009tc}.
This twist operator gives rise to a deformed statistics 
%on $\kappa$-Minkowski spacetime 
\cite{Young:2007ag,Govindarajan:2008qa}.
%However, since the $\kappa$-Poincar\'{e} Hopf algebra is a quantum symmetry
%described only approximatively by twisted quantum algebra,  
%there could be problems with identifying 
%charges and current conservation, that is with establishing the Noether theorem. 
%Thus, certain modification towards momentum conservation
%is necessary to obtain any reasonable physics out of approximatively
%twisted Hopf algebra prescription of our theory. 

Transformation from noncommutative $\kappa$-Minkowski to Minkowski spacetime 
in the case of the free field theory was described in \cite{Freidel:2006gc},
while the star product and interacting fields on $\kappa$-Minkowski space 
were treated in the same approach in \cite{KowalskiGlikman:2009zu}.
%The problem of UV/IR mixing and $\kappa$-deformation was discussed in \cite{Grosse:2005iz}.
%In correspondence to the above observations, 
%as well as by comparing deformed dispersion
%relations to corresponding time delay calculations of high energy photons,
%bounds can be put on the quantum gravity scale \cite{Borowiec:2009ty}.
We are continuing along the line where the main aim is to transcribe original NCQFT on 
$\kappa$-Minkowski to a corresponding commutative QFT on the standard Minkowski spacetime.
With this in mind,
we are considering the $\kappa$-deformation of a Minkowski spacetime whose
symmetry has a undeformed Lorentz sector 
(classical basis \cite{Kosinski:1994br,Borowiec:2009vb}). The noncommutative 
coordinates close in a $\kappa$-deformed Lie algebra and additionally, form a Lie
algebra together with the Lorentz generators
\cite{Meljanac:2006ui,Meljanac:2007xb,KresicJuric:2007nh,Meljanac:2010ps}.
This  deformation of the spacetime
structure affects the algebra of physical fields, leading to a
modification of multiplication in the corresponding universal
enveloping algebra, requiring the replacement of the usual pointwise multiplication by a
deformed star product, i.e. by the new star product ${\star}_h$.  
%thus reproducing important trace-like property \cite{Meljanac:2010ps}.
%The integral measure problems in the $\kappa$-Minkowski NC spacetime 
%are avoided since the measure function is naturally absorbed 
%within the new ${\star}_h$-product.  
%While computing the tadpole diagram contributions 
%to the scalar field propagation/self-energy, we anticipate
%that statistics on $\kappa$-Minkowski spacetime is specifically deformed. 
Next, the action is modified by truncating the initial model via expansion up to 
first order in the deformation parameter $a$, 
producing an effective model on commutative spacetime.
%We obtain further modified equations of motion 
%and conserved currents at that order,
%due to invariance under internal symmetry. 
%Truncating of the model was necessary to be able 
%to compute any relevant physical quantity, 
%such as self-energy of our complex scalar field $\phi$. 
%Above properties are very welcome, however 

We have to stress that by such a truncation of the $\kappa$-deformed action
%to the linear order in deformation parameter $a$ 
we have lost the nonperturbative quantum effects 
like the celebrated UV/IR mixing \cite{Grosse:2005iz}, 
which, amongst other, connects the noncommutative field theories 
with Holography \cite{Horvat:2010km,Cohen:1998zx} via UV and IR cutoffs, 
in a model independent way \cite{Horvat:2010km}.
Resummation of the expanded action could in principle restore the 
nonperturbative character of the theory, like the model in \cite{Horvat:2011iv}, 
thus producing UV/IR mixing \cite{Horvat:2011bs}, 
which under such circumstances would help
in determining what the UV theory might be. That is, it would help to
determine the UV completeness of the theory. 
However, we have to admit that for our $\kappa$-Minkowski $\phi^4$ model 
it is absolutely not clear that a resummation of higher orders in $a_\mu$ will give
a meaningful model beyond the tree level. 
Those are general properties of the most of NCQFT expanded/resummed in terms of 
the noncommutative deformation parameter. 
Holography and UV/IR mixing are in the literature known as
possible windows to quantum gravity 
\cite{Horvat:2010km,Cohen:1998zx,Szabo:2009tn}.

Our approach generally represents a {\it hybrid} approach modeling
between standard quantum field theory and NCQFT on 
$\kappa$-Minkowski spacetime involving $\kappa$-deformed momentum conservation law
\cite{Kosinski:1999ix,AmelinoCamelia:2001fd}.
%\cite{Kosinski:1999dw,Daszkiewicz:2004xy,KowalskiGlikman:2004qa,AmelinoCamelia:2001me}
%The results are discussed in the framework of two-point Green's function
%for low, middle, and for Planck scale energy regimes. We have found 
%semiclassical behavior of the first order quantum effects, and, as a consequence,
%that the mass term of the scalar field is shifted and these shifts are 
%very much different in low and high propagation energy regimes, respectively.
%Our $\kappa$-deformed dispersion relations are direction dependent and 
%we have found birefringence, \cite{Abel:2006wj,Buric:2010wd},
%of the massive scalar field mode. % at first order in deformation parameter $a$. 
%This is similar to the fermion field
%birefringence in truncated Moyal $\star$-product theories \cite{Buric:2010wd}.  

%The above describes the main results of this paper,
%which could be of physical importance, for example
%for the $\kappa$-NC scalar field (Higgs) and its deformed propagation.
%as well as to quantum gravity \cite{Maggiore:1993rv,Maggiore:1993zu,Giddings:2007bw}.

In the first section, we give some mathematical preliminaries including the
Hopf algebra structure of the $\kappa$-deformed Minkowski spacetime 
and the star products. In the second section, we introduce
the hermitian realization and the $\kappa$-deformed star product $\star_h$ 
corresponding to this realization.
The modified $\kappa$-deformed scalar field action based on the above notions 
is introduced next, and the equations of motion are derived, 
with the corresponding conserved currents. The properties of the 
$\kappa$-deformed action 
%in terms of the deformed model (with the undeformed fields) 
%on the ordinary Minkowski spacetime, 
and the perturbative study of the model two-point
Green's function, are discussed in the last section.
%That include the $\kappa$-deformed Feynman rules and the field propagation, 
%via computation of two-point Green's function, within the proposed model.    

\section{Mathematical preliminaries of $\kappa$-deformed Minkowski spacetime}
\subsection{$\kappa$-deformed algebra}

We are considering $\kappa$-deformation of Minkowski spacetime whose
symmetry has an undeformed Lorentz sector and whose noncommutative
coordinates $\x_{\mu}, ~ (\mu = 0,1,...,n-1),$ 
close in a Lie algebra together with the Lorentz generators
$M_{\mu\nu},~ (M_{\mu \nu} = -M_{\nu \mu}) $ \,,
\begin{align} 
\label{2.1}
 [\x_{\mu},\x_{\nu}] & = i(a_{\mu}\x_{\nu}-a_{\nu}\x_{\mu})\,, \\ 
 [M_{\mu \nu}, M_{\lambda \rho}]  & =  \eta_{\nu \lambda}M_{\mu \rho} -
 \eta_{\mu \lambda}M_{\nu \rho}
 -\eta_{\nu \rho} M_{\mu \lambda} + \eta_{\mu \rho} M_{\nu \lambda}\,, 
 \label{2.2} \\
 [M_{\mu\nu}, \x_{\lambda}] & = \x_{\mu} \eta_{\nu\lambda} - \x_{\nu}
  \eta_{\mu\lambda}-i\left( a_{\mu}  M_{\nu\lambda}-a_{\nu} 
  M_{\mu\lambda} \right)\,, 
  \label{2.3}
\end{align}
where the deformation parameter $a_{\mu}$ is a constant
Lorentz vector, and $~ \eta_{\mu\nu} = diag(-1,1,\cdot \cdot \cdot, 1)$
defines the metric in this spacetime. The quantity $a^2 = a_{\mu}a^{\mu}$ 
is Lorentz invariant having a dimension of inverse mass squared,
 $a^2 \equiv \frac{1}{{\kappa}^{2}}.$ 
  The above algebra has all the Jacobi identities satisfied,
 thus forming a Lie algebra with the property that in the
 limit $a_{\mu} \rightarrow 0,$ the commutative spacetime  with the
 usual action of the Lorentz algebra is recovered. 
 Throughout the paper we
shall work in units  $\hbar = c =1$. 

The symmetry of the deformed spacetime (\ref{2.1}) is assumed to be
described by an undeformed Poincar\'{e}
algebra. Thus, in addition to Lorentz generators  $M_{\mu\nu}, $ we
also introduce momenta $p_{\mu}$ which transform as vectors under the
Lorentz algebra, 
\begin{equation} 
[p_{\mu},p_{\nu}]=0\,,\;\;\;\;  
[M_{\mu\nu},p_{\lambda}]= \eta_{\nu\lambda}\,
p_{\mu}-\eta_{\mu\lambda}\, p_{\nu}\,. 
\label{2.4}
\end{equation}
For convenience, we refer to the algebra (\ref{2.1})-(\ref{2.5}) as the
deformed special relativity algebra since its different
realizations lead to different special relativity models with
different physics encoded in the deformed dispersion relations resulting
from such theories. This algebra, however, does not fix the commutation
relation between $~p_{\mu}$ and $\x_{\nu}$. In fact, there are
infinitely many possibilities for the commutation
relation between $~p_{\mu}$ and $\x_{\nu},$ all of which are consistent
with the algebra (\ref{2.1})-(\ref{2.5}) in the sense that the Jacobi
identities are satisfied between all generators of the algebra.   
In this way, we have an
extended algebra, which includes the generators $M_{\mu\nu}, ~p_{\mu}$ and $\x_{\lambda}$
and satisfies the Jacobi identities for all
combinations of the generators.
Particularly, the algebra generated by $p_{\mu}$ and $\x_{\nu}$ is a deformed
Heisenberg-Weyl algebra which in this paper we take to have the following form
%that can generally be written in the form
%\begin{equation} 
%\label{2.6}
%[p_{\mu},\x_{\nu}] = -i \Phi_{\mu\nu}(p)\,,
%\end{equation}
\begin{equation} 
[p_{\mu},\x_{\nu}]=-i\eta_{\mu\nu} \left(ap+\sqrt{1+ a^2 p^2}
  \right) +i a_\mu p_\nu\,.
\label{2.5}
\end{equation}

This particular type of phase space noncommutativity leads
to uncertainty relations of the form 
\cite{Maggiore:1993rv,Giddings:2007bw}
\begin{equation}
\triangle x_{\mu} ~ 
\ge ~ \frac{\hbar}{\triangle p_{\mu}}+\alpha G \triangle p_{\mu}\,,
\label{2.12}
\end{equation}
($\alpha$ is a constant and $G$ is the gravitational constant)
which have been obtained from the study of string collisions at
Planckian energies, i.e. so called gravity collapse of strings \cite{Giddings:2007bw}, 
thus manifesting its dynamical origin.
The same generalized uncertainty principle emerges from considerations
related to quantum gravity \cite{Maggiore:1993rv}.

The algebra (\ref{2.1})-(\ref{2.5}) can be realized by
\begin{eqnarray}
p_{\mu}&=&-i\partial_{\mu}=-i\frac{\partial}{\partial x^{\mu}},\;\;\;\;
M_{\mu\nu}=x_{\mu} \partial_{\nu}- x_{\nu} \partial_{\mu},
\nonumber\\
\x_{\mu}&=& x_{\mu} \sqrt{1+ a^2 p^2 } - iM_{\mu \nu} a^{\nu}\,,
\label{2.13}
\end{eqnarray}
where $\partial_{\mu}$ and $x_{\nu}$ are generators of the underformed Heisenberg algebra
$[x_{\mu},x_{\nu}]=[\partial_{\mu},\partial_{\nu}]=0$, $[\partial_{\mu},x_{\nu}]=\eta_{\mu\nu}$.
The phase space noncommutativity (\ref{2.5}), together with $\kappa$-Poincare algebra 
specified by relations (\ref{2.2}) and (\ref{2.4}) corresponds 
to the classical basis of the $\kappa$-Poincare algebra considered in 
\cite{Borowiec:2009vb,KowalskiGlikman:2002we,KowalskiGlikman:2002jr,Dimitrijevic:2003wv}.
The algebra (\ref{2.1})-(\ref{2.4}), as it is defined, is not complete. 
With the relations (\ref{2.5}), 
we obtain an extended algebra containing 
the deformed Heisenberg-Weyl as a subalgebra. 
%The realization (\ref{2.7}) indicates that deformed (\ref{2.1}),
%(\ref{2.4}), (\ref{2.6}) and undeformed (\ref{2.9}) 
%Heisenberg-Weyl algebras are isomorphic at the level of vector spaces.
%In the same way one can show that deformed and undeformed extended algebras
%are isomorphic in the same sence. It is of no importance 
%which concrete formula, (\ref{2.7}), is used for closing it.
%That was shown under some general setting in \cite{Borowiec:2010yw},
%where extended algebra (\ref{2.1})-(\ref{2.5}) 
%has been defined as crossed (smash) product algebra. 
In \cite{Borowiec:2010yw} the authors started from the coproduct 
(Hopf algebra + module algebra) and
then determined the crossed commutation relation of 
the extended algebra (\ref{2.1})-(\ref{2.5}). Our
route is just opposite: we close the algebra by the crossed commutation
relation and then accordingly determine the coproduct.

%%%%%%%%%%%%%%%%%%%%%%
\subsection{Hopf algebra and star product}
The symmetry underlying $\kappa$-deformed Minkowski space, characterized by the commutation relations
   (\ref{2.1}), is the deformed Poincar\'{e} symmetry which can most
 conveniently be described in terms of Hopf algebras. 
 As it was manifested in relations (\ref{2.2}) and (\ref{2.5}),
 the algebraic sector of this deformed symmetry is the same as that of the undeformed 
Poincar\'{e} algebra. However, 
%the action of the Poincar\'{e} generators
% on the deformed Minkowski space is modified in such a way, that the whole deformation
% is contained in the coalgebraic sector. This means that the Leibniz
% rules, which describe the action of the generators $M_{\mu\nu}$ and $p_{\mu}$ 
% on a product of fields,
%will no more have the standard form, but instead will be deformed
%and will depend on the $\Phi_{\mu\nu}$ realization. The Hopf algebra structure
% describes the properties of the generators of a deformed
%  Poincar\'{e} symmetry. Its algebraic
%  sector is determined by the relations (\ref{2.2}) and (\ref{2.5}).
%On the other hand, 
the coalgebraic sector is deformed and it is determined by the coproducts 
for translation ($p_{\mu}= -i\p_{\mu}$), rotation and the boost generators ($M_{\mu\nu}$)
\cite{Meljanac:2007xb,KresicJuric:2007nh},
\begin{eqnarray} 
\label{coproductmomentum}
\triangle \p_{\mu} &=& \p_{\mu}\otimes Z^{-1}+\mathbf{1}\otimes
\p_{\mu}+ia_{\mu} (\p_{\lambda} Z)\otimes
\p^{\lambda}-\frac{ia_{\mu}}{2} \square\, Z\otimes ia\p\,, \\
\triangle M_{\mu\nu} &=& M_{\mu\nu}\otimes
\mathbf{1}+\mathbf{1}\otimes M_{\mu\nu} \nonumber \\
&+& ia_{\mu}\left(\p^{\lambda}-\frac{ia^{\lambda}}{2}\square\right)\,
Z\otimes
M_{\lambda\nu}-ia_{\nu}\left(\p^{\lambda}-\frac{ia^{\lambda}}{2}\square\right)\,
Z\otimes M_{\lambda\mu} \,, 
\label{coproductangmomentum}
\end{eqnarray}
where $\otimes$ denotes the tensor product.
In the above expressions, $Z$ is the shift operator, determined by 
$Z^{-1}=ap+\sqrt{1+a^2p^2}$, $Z=1/Z^{-1}$. 
%and which itself has a simple expression for coproduct, $\triangle Z = Z\otimes Z$\,.
The operator $\square= \frac{2}{a^2}(1- \sqrt{1- a^2 \p^2})\,,$ 
is a deformed d'Alembertian operator 
\cite{Meljanac:2006ui,KresicJuric:2007nh},
which in the limit $a \rightarrow 0$ acquires the
standard form, $\square \rightarrow \p^2$, valid in undeformed
Minkowski space.

The Hopf algebra in question also has well defined counits and
antipodes. The antipodes for  the generators of the $\kappa$-Poincar\'{e} Hopf algebra 
and the related operator $Z$ are given by
\begin{equation} 
\label{3.8}
 S(\p_{\mu}) = \left( -\p_{\mu} + i a_{\mu} {\p}^2 +
 \frac{1}{2}a_{\mu}(a\p)\square \right) Z\,,
\end{equation}
\begin{equation} 
\label{3.8b}
 S(M_{\mu \nu}) = -M_{\mu \nu}
   + ia_{\mu} \left ( \p_{\alpha} - \frac{ia_{\alpha}}{2} \square
   \right ) M_{\alpha \nu}
  - ia_{\nu} \left ( \p_{\alpha} - \frac{ia_{\alpha}}{2} \square 
   \right ) M_{\alpha \mu}\,,
\end{equation}
while the counits remain trivial.
%In the above relations, the operator $Z$ is given by
%\begin{equation} 
%\label{3.8a}
%   Z \equiv \frac{1}{Z^{-1}} = \frac{1}{ -ia\p + \sqrt{1 - a^2 {\p}^{2}}}\,,    
%\end{equation}
%in accordance with $Z^{-1}$ and $\square$ as given above.

%Since we are interested in perturbative expansion of the field
%theoretic action,
%for later convenience we give $\triangle \p_{\mu} $ in form of a series
%expansion up to second order in the deformation parameter $a$,
%\begin{eqnarray} 
%  \triangle \p_{\mu} &=&  \p_{\mu}\otimes \mathbf{1} +\mathbf{1}\otimes
% \p_{\mu} 
%  - i\p_{\mu} \otimes a\p +
% ia_{\mu} \p_{\alpha} \otimes \p^{\alpha}  
% \nonumber \\
% & -& \frac{1}{2}a^2\p_{\mu}\otimes \p^2 
%  - a_{\mu} (a\p)\p_{\alpha} \otimes \p^{\alpha}
%   +\frac{1}{2} a_{\mu} \p^2 \otimes a\p
%    + {\mathcal{O}}(a^3)\,. 
%  \label{Dd}
%\end{eqnarray} 

Once we have the coproduct (\ref{coproductmomentum}), we can straightforwardly 
 construct a star product between two arbitrary
fields $f$ and $g$ of commuting coordinates \cite{Meljanac:2006ui,Meljanac:2007xb}.
 For the noncommutative spacetime (\ref{2.1}), the star product has the following form
\begin{equation} 
\label{d88}
(f \; \star \; g)(x)  =   \lim_{\substack{u \rightarrow x  \\ y \rightarrow x }}
 {\cal M} \left ( e^{x^{\mu} ( \triangle - {\triangle}_{0}) {\partial}_{\mu} }
    f(u) \otimes g(y) \right )\,, 
\end{equation}
where $ {\triangle}_{0}{\partial}_{\mu} =
 {\partial}_{\mu} \otimes 1 + 1 \otimes {\partial}_{\mu}$,
$ \: \triangle ({\partial}_{\mu})$ is given in (\ref{coproductmomentum})
and $ \; {\cal M} \; $ is the multiplication map in the undeformed Hopf algebra,
 namely, ${\cal M} (f(x) \otimes g(x)) = f(x) g(x)$ \cite{majid1}.
%From (\ref{d88}) we see that star product depends only on the
% coproduct for translation generators. 
%Its form 
% does not depend on the $\Phi_{\mu \nu}$ realization in
% (\ref{2.11}). However, since coproducts depend on the $\Phi_{\mu \nu}$
% realization, so does the star product according to (\ref{d88}), 
% in an implicit form. 
At this point we emphasize here once
 again that in this paper we are doing analysis based on the specific
 realization $\x_{\mu}= x_{\mu} \sqrt{1+ a^2 p^2 } - iM_{\mu \nu} a^{\nu}$ 
 and its hermitian variant (\ref{2.10d}), defined in the next section.
 The coproducts (\ref{coproductmomentum}) and (\ref{coproductangmomentum})
correspond to this particular type of realization. One can check
 that the star product (\ref{d88}) together with the coproduct
 (\ref{coproductmomentum}) is associative.
 
Note that the commutator (\ref{2.1}) can be written in terms of 
ordinary coordinates and the star product 
%$\phi (x) \star \psi(x)  ~ = ~ \hat{\phi}(\x) \hat{\psi}(\x) \triangleright 1 
%   =  ~ \hat{\phi}(\x) \triangleright (\hat{\psi}(\x)\triangleright 1)
%  ~ = ~ \hat{\phi}(\x)\triangleright \psi(x)$ 
from \cite{Meljanac:2010ps} as
\begin{eqnarray} 
\label{starcommutator}
[x_\mu,x_\nu]_{\star}=x_\mu \star x_\nu - x_\nu \star x_\mu = i(a_\mu x_\nu - a_\nu x_\mu)\,.
\end{eqnarray}
 
\section{Modified $\kappa$-deformed scalar field action}
%\subsection{Nonhermitian realization of the NC $\phi^4$ action}

In this section, we present an interacting scalar field model on
 noncommutative spacetime whose short distance geometry is governed by
 the $\kappa$-deformed symplectic structure (\ref{2.1}).
In order to obtain the physical meaning of the NC $\phi^4$ field theory, we have to 
introduce a complex scalar field $\phi$ with the accompanying notion of the
hermitian conjugation operation \cite{Meljanac:2010ps}.

\subsection{Hermitian realization of the NC $\phi^4$ action}

In order to obtain the hermitian action, we are necessarily forced to work with 
a hermitian realization represented by the operator ${\x}_{\mu}^{h}$, having the property
${({\x}_{\mu}^{h})}^{\ddagger} = {\x}_{\mu}^{h}.$ 
The hermitian operator $ {\x}_{\mu}^{h}$ can be constructed from the operator (\ref{2.13}) as
%\begin{equation} 
${\x}_{\mu}^{h}=\frac{1}{2}(\x_{\mu} + \x_{\mu}^{\ddagger})={({\x}_{\mu}^{h})}^{\ddagger}$\,,
%\label{2.10H}
%\end{equation}
which results in ($\ddagger$ here means the usual hermitian conjugation operation, 
$x_{\mu}^\ddagger=x_{\mu},\;\partial_{\mu}^\ddagger=-\partial_{\mu}$)
\begin{equation} 
\x_{\mu}^{h} = 
 x_{\mu} \sqrt{1+ a^2 p^2 } - iM_{\mu \nu} a^{\nu} - 
 i \frac{a^2}{2} \frac{1}{\sqrt{1+ a^2 p^2 }} p_{\mu}\,.
\label{2.10d}
\end{equation}
The change of the specific realization 
%into (\ref{2.10d}) in accordance with star product definition,
%(\ref{starproductdefinition}), 
%modifies the form of the star
%product and we obtain a the new star product denoted as ${\star}_h$:
% \begin{eqnarray} 
% \label{starproductdefinitionher}
% \phi(x){\star}_h\psi(x)~ =~\hat{\phi}({\x}^{h})\hat{\psi}({\x}^{h})\triangleright 1 
 %\nonumber \\
%   =  ~ \hat{\phi}({\x}^{h}) \triangleright (\hat{\psi}({\x}^{h}) \triangleright 1)
%  ~ = ~ \hat{\phi}({\x}^{h})\triangleright \psi(x)\,,
%\end{eqnarray}
%with ${\x}^{h}$ being given by (\ref{2.10d}).  Thus, we are 
forces us to replace the star product (\ref{d88}) with
a new one corresponding to the hermitian realization of 
the $\kappa$-Minkowski spacetime \cite{Meljanac:2010ps}:
\begin{equation} 
\label{newstarproduct}
(f \; {\star}_h \; g)(x)  =   \lim_{\substack{u \rightarrow x  \\ y \rightarrow x }}
 {\cal M} \left ( e^{x^{\mu} ( \triangle - {\triangle}_{0}) {\partial}_{\mu} }
  \sqrt[4]{\frac{1- a^2
   \triangle ( {\p}^2)}{(1- a^2 ~ {\p}^2 \otimes 1) ~ (1-
   a^2 ~ 1 \otimes {\p}^2 )}} ~
  f(u) \otimes g(y) \right )\,, 
\end{equation}
where it is understood that the coproduct $\triangle ({\p}_{\mu})$,
 Eq.(\ref{coproductmomentum}), is a homomorphism, i.e.
 $\triangle ({\p}^2) = \triangle ({\p}_{\mu}) \triangle ({\p}^{\mu})$.
In this way, the nonhermitian version of the star product (\ref{d88}) is
replaced by the above hermitian one.

The $\kappa$-deformed star product ${\star}_h$ (\ref{newstarproduct}) 
is associative in the same sense
as the star product (\ref{d88}). However the star product ${\star}_h$, 
contrary to the star product (\ref{d88}), has the 
same trace and/or integral property as the usual Moyal-Weyl product:
\begin{eqnarray} 
\int d^n x \;\phi^{\dagger} {\star}_h \psi = \int d^n x \;\phi^{*} \cdot \psi\,,
\label{property}
\end{eqnarray}
where the asterisk $*$ denotes usual complex conjugation,
and $\dagger$ means complex conjugation on the $\kappa$-Minkowski \cite{Meljanac:2010ps}.

The above results -- the new ${\star}_h$-product (\ref{newstarproduct}), and 
the identity (\ref{property}) -- embrace a very nice and important property:
the integral measure problems are avoided due to 
the measure function absorbtion 
within the new, $\kappa$-deformed, ${\star}_h$-product (\ref{newstarproduct}).  
In the new action 
\begin{eqnarray} 
\label{actionnew}
S_n[\phi] & = & \int d^n x ~ (\p_{\mu}\phi)^{\dagger} {\star}_h (\p^{\mu}\phi) + 
  m^2 \int d^n x ~ \phi^{\dagger} {\star}_h \phi    
\nonumber \\    
& + &\frac{\lambda}{4}  
\int d^n x~\frac{1}{2}(\phi^{\dagger}{\star}_h\phi^{\dagger}{\star}_h \phi {\star}_h \phi 
    +\phi^{\dagger} {\star}_h \phi {\star}_h \phi^{\dagger} {\star}_h \phi)\,,
\end{eqnarray}
the interaction $\phi^4$ term should in fact
incorporate six terms corresponding to all possible permutations of
fields $\phi$ and $\phi^{\dagger}.$ However, due to the integral property
(\ref{property}) of the star product (\ref{newstarproduct}), these six
permutations can be reduced to only two mutually nonequivalent terms.
% $\phi^{\dagger} {\star}_h \phi^{\dagger} {\star}_h \phi {\star}_h \phi$ and
% $\phi^{\dagger} {\star}_h \phi {\star}_h \phi^{\dagger} {\star}_h \phi$.

When expanded up to the first order in the deformation parameter $a$,
the action (\ref{actionnew}), after rearrangements including
integration by parts, receives the following form
\begin{eqnarray} 
  S_n[\phi] 
 & = & \int d^n x ~ \Big[(\p_{\mu}\phi^{*}) (\p^{\mu}\phi) +
 m^2\;\phi^{*}\phi + \frac{\lambda}{4}{(\phi^{*}\phi)}^2\Big] 
 \label{actioncomlex}\\
 & + & 
 i\frac{\lambda}{4}\int d^n x \bigg[ a_{\mu} ~ 
 x^{\mu} \Big({\phi^{*}}^2(\p_{\nu} \phi) {\p}^{\nu} \phi -  
 \phi^{2}(\p_{\nu} {\phi}^{*}) {\p}^{\nu} {\phi}^{*}\Big)  
\nonumber \\
 & + & 
 a_{\nu} ~ x^{\mu} \Big(\phi^{2}(\p_{\mu} \phi^{*}) {\p}^{\nu} \phi^{*} -  
 {\phi^{*}}^2(\p_{\mu} {\phi}) {\p}^{\nu} {\phi}\Big)  
%  \nonumber \\
 % &  + &       
 +   \frac{1}{2}  a_{\nu} x^{\mu}~ \phi^{*}\phi \Big( 
    (\p_{\mu} {\phi}^{*}) {\p}^{\nu} \phi
     -(\p_{\mu} \phi) {\p}^{\nu} {\phi}^{*}\Big)  \bigg ]
 %   + {\mathcal{O}}(a^2)
    \, .
    \nonumber 
\end{eqnarray}
%Note that the oscillator term proportional to $\xi^2$ attain no correction 
%in the deformation parameter $a$. 
%These two features of our model separate completely in above action.

At the end of this subsection note that the Hopf algebra, yielding (\ref{actionnew}), 
is a twisted symmetry algebra, where existence/conservation of charges and currents
are still subject of research. 
However the action (\ref{actioncomlex}), obtained by expansion of (\ref{actionnew}) 
up to the first order in the deformation parameter $a$,
is invariant under the internal symmetry transformations.
%\begin{equation}
%\left(
%    \begin{array}{c}
%        \phi \\
%       \phi^{\ast}
%    \end{array}
%\right)
%\rightarrow
%%\end{equation}
%%\begin{equation}
%\begin{pmatrix}
%    e^{i\chi} &0 \\
%   0 & e^{-i\chi}
%\end{pmatrix}
%\left(
%    \begin{array}{c}
%        \phi \\
%       \phi^{\ast}
%    \end{array}
%\right)\,.
%\label{intsym}
%\end{equation}
%thus the corresponding Noether current should be conserved. 

\subsection{Equations of motion and Noether currents of internal symmetry}

We proceed further by determining the equations of motion for the fields
$\phi$ and ${\phi}^{\ast}$: 
\begin{eqnarray} 
\label{eom2}
( {\p}_{\mu} {\p}^{\mu} - m^2 )\phi  
 & = & \frac{\lambda}{6}  
 \Big[{\phi}^{\ast} {\phi}^2 + 
  ia_{\mu} x^{\mu}\Big( {\phi}^2 
 {\p}_{\nu} {\p}^{\nu} {\phi}^{\ast} + 
   {\phi}^{\ast}
 ({\p}_{\nu} \phi) {\p}^{\nu} \phi
  + \phi 
 ({\p}_{\nu} {\phi}) {\p}^{\nu} {\phi}^{\ast} \Big)  
 \nonumber\\
 & - & ia^{\mu}x^{\nu} \Big( {\phi}^{\ast} 
 ({\p}_{\nu} \phi) {\p}_{\mu} \phi 
 + \phi 
 ({\p}_{\nu} \phi) {\p}_{\mu} {\phi}^{\ast}
 +  {\phi}^2 
 {\p}_{\mu} {\p}_{\nu} {\phi}^{\ast} 
  +  \phi ({\p}_{\mu} \phi)  {\p}_{\nu} {\phi}^{\ast}\Big)  
 \nonumber \\    
 & + & \frac{i}{4} (1-n) a^{\mu} {\phi}^{\ast} \phi 
 {\p}_{\mu} \phi + \frac{i}{2} (1-n) a^{\mu} {\phi}^2 
 {\p}_{\mu} {\phi}^{\ast} \Big]\,,
\end{eqnarray}
where $\phi = 0$ is the trivial solution of the above equation, as it should be.
The equation of motion for ${\phi}^{\ast}$ can be obtained from (\ref{eom2}).
 
Next, we present Noether currents derived from 
the Lagrangian densities (\ref{actioncomlex}):
\begin{eqnarray}
j^{\mu}(x)&=&i\frac{\delta{\cal L}}{\delta(\p_\mu \phi)}\phi -
i\frac{\delta{\cal L}}{\delta(\p_\mu {\phi}^{\ast})}{\phi}^{\ast}\,,
\label{current}\\
\frac{\delta{\cal L}}{\delta(\p_\mu \phi)}
&=&
\frac{1}{2}\p^{\mu}{\phi}^{\ast}+ 
\frac{i\lambda}{4} 
\Big[{{\phi}^{\ast}}^2 (2a_{\nu}x^{\nu} \p^{\mu}-a^{\nu}x^{\mu}\p_{\nu}
-a^{\mu}x^{\nu}\p_{\nu})\phi +
\frac{1}{2} {\phi}^{\ast}\phi (a^{\mu}x^{\nu}\p_{\nu}-
a^{\nu}x^{\mu}\p_{\nu}){\phi}^{\ast} \Big]\,,
\nonumber\\
\frac{\delta{\cal L}}{\delta(\p_\mu {\phi}^{\ast})}
&=&
\frac{1}{2}\p^{\mu}{\phi} - 
\frac{i\lambda}{4} 
\Big[{\phi}^2 (2a_{\nu}x^{\nu} \p^{\mu}-a^{\nu}x^{\mu}\p_{\nu}
-a^{\mu}x^{\nu}\p_{\nu})\phi^{\ast} +
\frac{1}{2} {\phi}^{\ast}\phi (a^{\mu}x^{\nu}\p_{\nu}-
a^{\nu}x^{\mu}\p_{\nu}){\phi} \Big]\,.
\nonumber
\end{eqnarray}
The above current (\ref{current}), is conserved; that is $\p_{\mu}j^{\mu}(x)=0$\,, 
as it should be due to the invariance of the Lagrangian (\ref{actioncomlex}) 
under the internal symmetry transformations. This can be showed by 
straightforward computations from (\ref{actionnew})-(\ref{current}).

\section{Perturbative study of the model two-point function}

\subsection{Feynman rules}

Even though the S-matrix LSZ formalism, including Wick theorem, is not quite clearly 
defined on $\kappa$-Minkowski noncommutative spacetime, we continue bona fide towards the 
computation of the quantum properties of the model described by the action (\ref{actioncomlex}).
%To do that, we first derive relevant Feynman rules and than compute the tadpole diagram
%contributions to the 2-point Green's function of our model.

Due to the $\kappa$-deformation of our model,  
the statistics of particles is twisted, 
so that we are generally no more dealing with pure bosons.
We are in fact dealing with {\it something} whose statistics is governed by 
the statistics flip operator
\cite{Balachandran:2006pi,Balachandran:2010xk,Balachandran:2010wq} 
and the quasitriangular structure
(universal R-matrix) on the corresponding quantum group 
\cite{Borowiec:2010yw,vgdrinfeld,faddeev,Rmatrixmajid}. 
It would be interesting to investigate  
these mutual relations more thoroughly,
but at  the surface level, we can argue that it is possible 
to pick up the basic characteristics of the twisted statistics by using 
the nonabelian momentum addition law 
\cite{Freidel:2006gc,KowalskiGlikman:2009zu,KowalskiGlikman:2002ft,Lukierski:2002wf}. 
It can be seen that the accordingly induced deformation of the $\delta$-function 
(arising from the implementation of the nonabelian momentum
addition/subtraction rule) yields the usual $\delta$-function 
multiplied by a certain statistical factor
%which has its origin in the $\kappa$-modified statistics.
When we speak about deformed statistics, we have in mind a less rigid
notion of statistics as applied to the symmetry properties of the states,
where multiparticle change of the 4-momenta may change the state's symmetry properties.

In order to obtain the Feynman rules in momentum space, 
we are suggesting to use the following line of reasoning,
which we shall further on call {\it hybrid} approach. \\
(A) We use the methods of standard QFT and treat the modifications 
in action (\ref{actioncomlex}) as a perturbation. 
In doing this, we obtain propagators and Feynman rules for vertices. \\
(B) We know that 
the statistics of particles is twisted and that it has to 
be implemented into the formalism.
Thus, we require that the ordinary addition/subtraction rule induces
deformed rule for twisted statistics on the $\kappa$-Minkowski spacetime.
This, in momentum space, means 
\begin{equation}
\sum_i k^{\mu}_i\;\; \rightarrow \;\;
\sum_{\oplus i} k^{\mu}_i  \;\;\;\;\,\&\,\;\;\;\;
\sum_i k^{\mu}_i - \sum_j p^{\mu}_j\;\; \rightarrow \;\;
\sum_{\oplus i} k^{\mu}_i \ominus \sum_{\oplus j} p^{\mu}_j \,,
\label{oplus+}
\end{equation}
where induced {\it deformed addition/subtraction} rules are going to be
defined in Subsection 4.1.2, for the simplest cases of two to four momenta.   
The associativity of the the direct sum $\oplus$ 
is satisfied due to the associativity of the
star product (\ref{newstarproduct}).
We proceed in two steps:\\
(1) Following above arguments we implement 
the induced conservation law within the delta functions in 
the Feynman rule, and\\
(2) whenever needed, we use the modified/deformed conservation law along the course
of evaluation of the Feynman diagrams.

\subsubsection{Feynman rules (A): standard momentum addition law}

From now on we continue to work with Euclidean metric.
In transition from Minkowski to Euclidean signature
we are using the transition rules: 
$a^M=(a^M_0,a^M_i) \longrightarrow a^E=(a^E_i, a^E_n)$,
where $a^E_n=ia_0^M$, and similary for any $n$-vector. Thus the scalar product
is defined as $a^E k^E=a^E_{\mu}k^E_{\mu}=a^E_i k^E_i + a^E_n k^E_n
=-a^M_0 k^M_0 + a^M_i k^M_i= a^M k^M$. 
In subsequent consideration we drop the $M$/$E$ superscripts, 
but it is understood that we work with Euclidean quantities.

In the further computation we are using the following full propagator 
\begin{equation}
G\equiv G(k_1,k_2) = \frac{i}{k_1^2+m^2}\delta^{(n)}(k_1-k_2)\,.
 \label{propagator}
\end{equation}
%This is going to be used .
%We believe that the approximative expression (\ref{propagator1}) is good enough
%to help us to indicate the influence of 
%the $\xi^2$ term on the one-loop quantum corrections.
%This is going to be presented in Subsection 4.2.
%Of course the full computation of quantum corrections including 
%the Mehler kernel is out of
%scope for this paper, but it is certainly going to be performed in the future.

%If $a$ is of the order of the Planck length, $\xi$
%despite being small, carries contributions to Green's functions
%that are  still larger than the terms linear in $a$.

The vertex function, illustrated in Fig.~\ref{fig:vertex}, in the momentum space is given by
\begin{equation}
\tilde\Gamma (k_1,k_2,k_3,k_4;a)=i\frac{\delta^4 S[\tilde \phi]}{\delta\tilde\phi(k_1) 
\delta\tilde\phi(k_2)\delta\tilde\phi^*(k_3) \delta \tilde \phi^*(k_4)} \,,
\label{vertex}
\end{equation} 
and amounts to the following expression:
\begin{figure}
\begin{center}
\includegraphics[width=40mm]{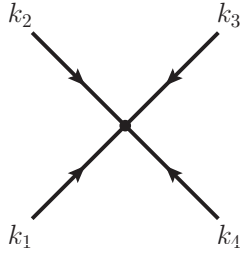}
\end{center}
\caption{Scalar 4-field vertex}
\label{fig:vertex}
\end{figure}

\begin{eqnarray}
\nonumber
 \tilde\Gamma(k_1,k_2,k_3,k_4;a) &= &i(2\pi)^n\frac{\lambda}{2}\,a_{\nu} 
\Bigg[\frac{a_\nu}{a^2}+ 
\frac{1}{4}\bigg(k_{4\mu} k_{3\nu}+k_{3\mu}k_{4\nu}-2\delta_{\mu\nu}k_{4\rho}k_{3\rho}\\
\nonumber
&&\hspace{-2.5cm}
+\frac{1}{2}(k_{2\mu}k_{4\nu}-k_{4\mu}k_{2\nu}+k_{2\mu}k_{3\nu}-k_{3\mu}k_{2\nu})\bigg)
\partial_\mu^{k_1}\\
\nonumber
&& \hspace{-3cm}
+ \frac 14 \bigg( k_{4\mu} k_{3 \nu} + k_{3\mu} k_{4\nu} 
- 2 \delta_{\mu\nu} k_{4\rho} k_{3\rho} \\
\nonumber
&& \hspace{-2.5cm}
+ \frac 12 ( k_{1\mu}k_{4\nu} - k_{4\mu}k_{1\nu} + k_{1\mu}k_{3\nu} - k_{3\mu}k_{1\nu} )
\bigg) \partial_\mu^{k_2}\\
%%
%\nonumber
&& \hspace{-3cm}
+ \frac 14 \bigg(
k_{1\mu}k_{2\nu} + k_{2\mu}k_{1\nu} - 2 \delta_{\mu\nu} k_{1\rho}k_{2\rho}
\bigg) ( \partial_\mu^{k_3} + \partial_\mu^{k_4} )
\Bigg] \delta^{(n)} (k_1 + k_2 - k_3 - k_4),
\label{Feynrule}
\end{eqnarray}
where we denote $\partial_\mu^k = \frac{\partial}{\partial k_\mu}$, 
and all four momenta $k_i$ are flowing into the vertex.
The coupling $\lambda$ has to be dimensionally regularized.

\subsubsection{Feynman rules (B): $\kappa$-deformed momentum addition law}

Next, we discuss the notion which anticipates the
induced momentum conservation law on the $\kappa$-space, 
within our {\it hybrid} approach. 
Namely, the $\delta$-function in (\ref{Feynrule})  
comes from the contraction of fields, where 
the momentum conservation should be obeyed in accordance with
the $\kappa$-deformed momentum addition rule \cite{Freidel:2006gc}.
We have two cases for summation/subtraction of 4-vectors $k^{\mu}$
with respect to the physical situation of four particles and/or quantum fields
propagating in space with respect to an interaction point:\\
(I) all particle momenta flowing into the vertex, as given in Fig. \ref{fig:vertex}
\begin{equation}
k_{1\mu} + k_{2\mu} + k_{3\mu} + k_{4\mu} =0\;\; \to \;\;
k_{1\mu} \oplus k_{2\mu} \oplus k_{3\mu} \oplus k_{4\mu} =0\,,
\label{oplus}
\end{equation}
(II) the process of scattering "2 particle $\to$ 2 particle", where we have 
\begin{equation}
(k_{1\mu} + k_{2\mu}) - (k_{3\mu} + k_{4\mu}) =0\,
\;\; \to \;\;
(k_{1\mu} \oplus k_{2\mu}) \ominus (k_{3\mu} \oplus k_{4\mu}) =0\,.
\label{ominus}
\end{equation}

Having defined the Feynman rules (\ref{propagator}) and (\ref{Feynrule}),
we have completed the first stage of our program, that is to deduce
the free propagation and interaction properties of the model by using the standard
quantization.
 
At this point we turn to the second part, which 
includes the effective description of the statistics of  particles
described by the model. As already indicated before, the statistics of
particles is twisted in $\kappa$-space 
\cite{Govindarajan:2009wt,Govindarajan:2008qa}, with the deformation 
being encoded in the nonabelian momentum addition rule. It is
known that the rule for addition of momenta is governed by the
coproduct structure of the Hopf algebra in question. In our case,
the relevant Hopf algebra is the $\kappa$-Poincar\'{e} algebra and the
corresponding coalgebra structure is given by (\ref{coproductmomentum}),
 (\ref{coproductangmomentum}), and (\ref{3.8}), (\ref{3.8b}).
In particular, the coproduct (\ref{coproductmomentum}) for translation
generators determines the required momentum addition rule,  
which in the momentum space and up to the first order in
deformation $a$, from (\ref{3.8}) and by the expansion
$  \triangle \p_{\mu} =  \p_{\mu}\otimes \mathbf{1} +\mathbf{1}\otimes \p_{\mu} 
  - i\p_{\mu} \otimes a\p +
 ia_{\mu} \p_{\alpha} \otimes \p^{\alpha}  
  - \frac{1}{2}a^2\p_{\mu}\otimes \p^2 
  - a_{\mu} (a\p)\p_{\alpha} \otimes \p^{\alpha}
   +\frac{1}{2} a_{\mu} \p^2 \otimes a\p
    + {\mathcal{O}}(a^3)\,$, yields: 
\begin{eqnarray}
 S(p_{\mu}) &=&-iS(\partial_{\mu})=-p_{\mu}-a_{\mu}p^2+(ap)p_{\mu} 
 +{\mathcal O}(a^2)\,,
\label{4} \\
(p_{\mu} \oplus k_{\mu}) &=& (p+k)_{\mu} + (ak)p_{\mu} - a_{\mu}(p k)
+{\mathcal O}(a^2)\,.
\label{4+}
\end{eqnarray}
Here we have the nonabelian momentum addition rule (\ref{4+}), while
$S(p)$ is the antipode with the property 
$p^{\mu} \oplus S(p^{\mu}) = 0$, which in fact represents the very definition of 
the antipode. 
Namely, since commutativity in momentum space is not satisfied, 
i.e. $k\oplus p \not= p\oplus k$, a certain ordering has to be implemented.
However, instead of implementation of a possibly complicated unknown ordering,
we shall proceed in the most simple way by taking into account all possible types 
of contributions. This is symmetric ordering; 
for example $k\oplus p\oplus q$, $p\oplus k \oplus q$, etc.
Combining (\ref{4}) and (\ref{4+}) 
we obtain the following momentum subtraction rule:
\begin{eqnarray}
p_{\mu} \ominus k_{\mu}  &\equiv& (p \oplus S(k))_{\mu} = 
 (p-k)_{\mu}(1 - ak) + a_{\mu}(p k - k^2)
 +{\mathcal O}(a^2)\,.
\label{5}
\end{eqnarray}
This enables us to rewrite the energy-momentum conservation which is
assumed to be satisfied at each vertex. Thus, if two external momenta
$k_1$ and $k_2$ flow into the vertex and the other two external
momenta $k_3$ and $k_4$ flow out of the vertex, then, written in components, we have
the induced momentum conservation law (\ref{ominus}).
%which corresponds to our physical situation while computing the 4-field
%tadpole diagram in the next subsection.

In order to obtain the expressions for the $\delta$-functions in 
the Feynman rules, we start with
\begin{equation}
\delta^{(n)}(p \ominus k) = \sum_i
\Bigg|\det \Bigg(\frac{\partial {(p \ominus k)}_{\mu}}{\partial p_{\nu}}\Bigg)_{p = q_i} 
\Bigg|^{-1}
\delta^{(n)}(p-q_i) \,,
\label{6}\end{equation}
where we have to sum over all zeros $q_i$ for 
the expression in the argument of the $\delta$-function on the LHS. 
Since there is only one zero, $q_i=k$, with the help of subtraction rule (\ref{5}),
we find the following first order contribution to the above
$\delta^{(n)}$-function
\begin{equation}
\delta^{(n)}(p \ominus k) = \frac{\delta^{(n)}(p - k)}{{(1 - ap)}^{n-1}} = 
   (1 + (n-1)ap + \mathcal{O}(a^2))\delta^{(n)}(p -k) \,.
\label{7}
\end{equation}

It was shown in \cite{Meljanac:2010ps} that the star product $\star_h$ 
(\ref{newstarproduct}) breaks translation invariance
(in the sense of the definition introduced in \cite{Kosinski:1999dw}). 
However, this feature does not show up 
until the computations are extended to second order in
the deformation parameter $a$.
The important point is that translation invariance is intact 
to first order. Since we are carrying out our study
in exactly this order, we are allowed to invoke the energy momentum conservation
albeit in a modified form, dictated by the modified 
coproduct structure. The relation between Hopf algebra symmetries
and conservation laws is an important subject of investigation. 
This is the issue of generalizing the Noether theorem
%thus the whole subject is still appealing 
\cite{Agostini:2006nc}.

With the idea of implementing the new physical features that have just been described,
we modify the Feynman rule (\ref{Feynrule}).
With the help of (\ref{4+}), (\ref{5}), and,
in the spirit of our {\it hybrid} approach \cite{Kosinski:1999ix,Kosinski:2003xx},
by choosing the following replacement of the $\delta$-function in (\ref{Feynrule})
\begin{equation}
\delta^{(n)} (k_1 + k_2 - k_3 - k_4) \rightarrow
\delta^{(n)} ((k_1 \oplus k_2) \ominus (k_3 \oplus k_4))
+\delta^{(n)} ((k_1 \oplus k_2) \ominus (k_4 \oplus k_3))\,,
\label{delta}
\end{equation}
we obtain the {\it hybrid} Feynman rule which obeys the
$\kappa$-deformed momentum addition/subtraction rule
using the sum of the $\delta$-functions (\ref{delta}):
\begin{eqnarray}
\nonumber
 \tilde\Gamma(k_1,k_2,k_3,k_4;a) &= &i(2\pi)^n\frac{\lambda}{2}\,a_{\nu} 
\Bigg[\frac{a_\nu}{a^2}+ 
\frac{1}{4}\bigg(k_{4\mu} k_{3\nu}+k_{3\mu}k_{4\nu}-2\delta_{\mu\nu}k_{4\rho}k_{3\rho}\\
\nonumber
&&\hspace{-2.5cm}
+\frac{1}{2}(k_{2\mu}k_{4\nu}-k_{4\mu}k_{2\nu}+k_{2\mu}k_{3\nu}-k_{3\mu}k_{2\nu})\bigg)
\partial_\mu^{k_1}\\
\nonumber
&& \hspace{-3cm}
+ \frac 14 \bigg( k_{4\mu} k_{3 \nu} + k_{3\mu} k_{4\nu} 
- 2 \delta_{\mu\nu} k_{4\rho} k_{3\rho} \\
\nonumber
&& \hspace{-2.5cm}
+ \frac 12 ( k_{1\mu}k_{4\nu} - k_{4\mu}k_{1\nu} + k_{1\mu}k_{3\nu} - k_{3\mu}k_{1\nu} )
\bigg) \partial_\mu^{k_2}\\
\nonumber
&& \hspace{-3cm}
+ \frac 14 \bigg(
k_{1\mu}k_{2\nu} + k_{2\mu}k_{1\nu} - 2 \delta_{\mu\nu} k_{1\rho}k_{2\rho}
\bigg) ( \partial_\mu^{k_3} + \partial_\mu^{k_4} )\Bigg]\\
&& \hspace{-2.5cm}\times \bigg[
\delta^{(n)} ((k_1 \oplus k_2) \ominus (k_3 \oplus k_4))
+\delta^{(n)} ((k_1 \oplus k_2) \ominus (k_4 \oplus k_3))
\bigg]\,.
\label{Feynrulekappa}
\end{eqnarray}
In the above, the sum of the $\delta$-functions
represents all mutually different physical situations.

The $\delta$-functions in (\ref{Feynrulekappa}), should in principle  
come from the contraction of fields quite naturally, if the
noncommutative version of the LSZ formalism is applied to our
model. Since such a formalism has not been developed so far, we choose to
follow a kind of {\it hybrid} approach that combines the standard quantum field theory
consideration 
%(used when treating terms in the Lagrangian linear in $a$ as small perturbations) 
with the peculiarities  resulting from 
the statistics properties of particles in
$\kappa$-space. The latter part is realized through embedding a
nonabelian momentum-energy conservation within the 4-point vertex function. 
%That approach may seem as a {\it hybrid} construction raised in
%trying to move our understanding one step forward towards a
%complete quantum theory on noncommutative spaces in general.
In this sense, {\it the hybrid} approach can serve as an
intermediate step bridging the gap between the standard quantum field theory and the
complete field theory on $\kappa$-space. 
%in as much the similar way as
%for example the semiclassical theory of radiation can be considered as
%a cross-over towards the quantum theory of radiation.
%In following we have to 
%rely in part on intuition, especially  when peculiarities of $\kappa$-space
%statistics have to be taken into account. 

The Feynman rule (\ref{Feynrulekappa}) appears to be consistent
with the energy-momentum conservation that respects 
$\kappa$-deformed momentum addition rule.
In order to obtain the complete expression for 
the $\delta$-functions appearing in (\ref{Feynrulekappa}),   
we are proceeding in two steps.
With the help of (\ref{4+})/(\ref{7}), up to linear order in $a$,
and with $j,l=3,4;\, j\not= l$, we have:
\begin{eqnarray}
\delta^{(n)} ((k_1 \oplus k_2) \ominus (k_j \oplus k_l))
&=&\frac{\delta^{n}((k_1 \oplus k_2) - (k_j \oplus k_l))}
{{\bigg(1 - a(k_1 \oplus k_2)\bigg)}^{n-1}} 
\label{deltakappa}\\ 
&& \hspace{-1.3cm}
=\frac{\delta^{(n)}((k_1 \oplus k_2)-(k_j \oplus k_l))}
{{\bigg[1-\bigg(a(k_1 + k_2)+(ak_1)(ak_2)-a^2(k_1k_2)
+ {\mathcal O}(a^3)\bigg)\bigg]}^{n-1}}
\nonumber\\
&& \hspace{-1.3cm}
=(1+(n-1)a(k_1+k_2)+ {\mathcal O}(a^2))
\delta^{(n)}((k_1\oplus k_2)-(k_j \oplus k_l)) \,.
\nonumber
\end{eqnarray}
The second step is to compute the delta functions from (\ref{deltakappa})
in the same way as in (\ref{7}):
\begin{equation}
\delta^{(n)}((k_1\oplus k_2)-(k_j \oplus k_l))=\sum_i
\Bigg|\det \Bigg(\frac{\partial {((k_1 \oplus k_2) 
-(k_j \oplus k_l)}_{\mu}}{\partial k_{1\nu}}\Bigg)_{k_1=q_i} 
\Bigg|^{-1}
\delta^{(n)}(k_1-q_i) \,,
\label{6a}
\end{equation}
where we have to sum up over all zeros $q_i$ for 
the expression in the argument of the $\delta$-function. 

Next, we shall choose the specific momenta $k_2=k_3=\ell$
we need for the evaluation of the tadpole diagram.
Because there are no zeros for the delta function 
$\delta^{(n)}((k_1 \oplus \ell) - (\ell \oplus k_4))$, the only 
contribution comes from the second combination 
$\delta^{(n)}((k_1\oplus\ell)-(k_4\oplus\ell))$.
In order to perform that computation, 
we start with (\ref{4+}) and  orient the vector $a$
in the direction  of time, $a =(0,...,0,ia_0). $ 
Due to covariance, the obtained result will also
 be valid for an arbitrary orientation of $a$. Hence
\begin{eqnarray}
\label{Jacobian}
  \det {\left( \frac{\p {((k_{1} \oplus \ell) - (k_{4} \oplus
          \ell))}_{\mu}}{\p k_{1\nu }} \right)}_{k_{1} = k_{4}}
&=&  \left| \begin{array}{ccccc}
  1 & ~ - ia_0 \ell_1 & 
   \cdot \cdot \cdot & ~ - ia_0 \ell_{n-2}
   & ~  - ia_0 \ell_{n-1} \\
 0  & 1+ a\ell &  \cdot \cdot \cdot
 & 0 & 0 \\ 
 \cdot & \cdot &   \cdot \cdot \cdot & \cdot & \cdot \\
\cdot & \cdot &  \cdot \cdot \cdot & \cdot & \cdot \\
 0  &  0 &  \cdot \cdot \cdot 
 & 1 + a\ell & 0 \\
  0 & 0 & \cdot \cdot \cdot
 & 0 &  1 + a\ell \\
\end{array} \right|  
\nonumber\\
&=&  {(1 + a\ell)}^{n-1}.
\end{eqnarray}
Since there is only one zero, $q_i=k_4$, we find 
\begin{eqnarray}
\hspace{-5mm}
\delta^{(n)}((k_1 \oplus \ell) - (k_4 \oplus \ell))  
=\frac{\delta^{(n)}(k_1 - k_4)}{{(1 + a\ell)}^{n-1}}
= \bigg (1 - (n-1) a\ell  \bigg ) \delta^{(n)}(k_1 - k_4) 
+\mathcal O(a^2) \,,
\label{tadpoleaksi3}
\end{eqnarray}
which gives the final expression
\begin{eqnarray}
\delta^{(n)}((k_1 \oplus \ell) \ominus (k_4 \oplus \ell ))
&=&\bigg(1+(n-1)a(k_1+\ell)\bigg)\frac{\delta^{(n)}(k_1 -k_4)}{{(1+a\ell)}^{n-1}}
+\mathcal O(a^2)
\nonumber\\
&=&(1 + (n-1)ak_1)\delta^{(n)}(k_1  - k_4) + \mathcal O(a^2)\,.
\label{delta-}
\end{eqnarray}
In the above expression the $\ell$ dependences drop out as we
expected, thus showing the consistency of the {\it hybrid} Feynman rule derivation. 
The remaining factor $(1 + (n-1)ak_1)$ in (\ref{delta-}) 
is due to the $\kappa$-space twisted 
particle statistics of our {\it hybrid} approach.

\subsection{Scalar field propagation }
\subsubsection{Tadpole diagram: standard momentum conservation}

\begin{figure}
\begin{center}
\includegraphics[width=70mm]{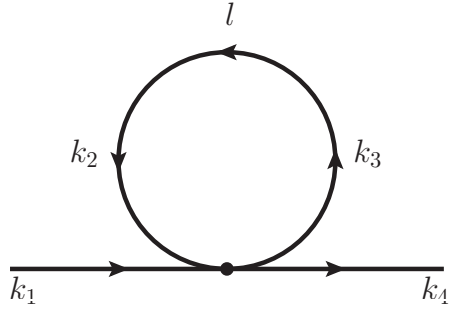}
\end{center}
\caption{Scalar 4-field tadpole}
\label{fig:tadpol}
\end{figure}

In order to compute the tadpole diagram from Fig. \ref{fig:tadpol},
using dimensional regularization,
we have to introduce in the action (\ref{actioncomlex}) 
new dimensionful regularization masses denoted by $\mu$
for the coupling $\lambda$.
In accordance with QFT \cite{Casalbuoni}, the
regularization of the $\phi^4$ model requires:
\begin{eqnarray}
\lambda_{new}&=&\lambda_{old}(\mu^2)^{\frac{n}{2}-2}
\;\;\; \to \;\;\;(\mu^2)^{2-\frac{n}{2}}\lambda\,,\;\;\;\lambda=\lambda_{new}\,.
\label{lambdamu}
%\\
%\xi_{new}&=&\xi_{old}({\mu'}^2)^{\frac{n}{2}-4}
%\;\;\; \to \;\;\;({\mu'}^2)^{4-\frac{n}{2}}\xi\,,\;\;\;\xi=\xi_{new}\,.
%\label{ximu}
\end{eqnarray}
%Here $\mu'$ is defined in a way to retain the same dimension of 
%constant $dim[\xi^2]=dim[mass^4]$, for $n=4$, 
%under the loop integral contribution from $\xi$-term in (\ref{actioncomlex}).

We shall further restrict the computation only to the
first order contribution of the two-point function ${\Pi}^{a}_2$ in our 
model (\ref{actioncomlex}) corresponding to the diagram from Fig. \ref{fig:tadpol}:
\begin{equation}
\Pi^{a}_2 = 
{\Pi}^{0}_2+{\Pi}^{a\ne 0}_2
%+{\Pi}^{0,\xi\ne 0}_2+{\Pi}^{a\ne 0,\xi\ne 0}_2
=
\int\frac{d^n k_2}{(2\pi)^n}\frac{d^n k_3}{(2\pi)^n}\,
\tilde\Gamma(k_1,k_2,k_3,k_4;a,\mu) G(k_2,k_3;\mu)\,.
\label{3}
\end{equation} 
In general, the one-loop integral (\ref{3}) produces nonzero contributions, 
where $G(k_2,k_3;\mu)$ is given by (\ref{propagator}). 
%further on. 
%In computing those two terms from (\ref{3}) we 
Assuming momentum conservation, $k_1 + k_2 = k_3 + k_4$ and $k_2 = k_3 = \ell$ we have
\begin{eqnarray}
\hspace{-.5cm}
{\Pi}^{a}_2=
%{\Pi}^{0,0}_2+{\Pi}^{a\ne 0,0}_2=
%+{\Pi}^{0,\xi\ne 0}_2+{\Pi}^{a\ne 0,\xi\ne 0}_2=  
%T^{0,0}_2+T^{a,0}_2+T^{0,\xi}_2=
\int \frac{d^n\ell}{(2\pi)^n} \;
\tilde\Gamma(k_1,\ell,\ell,k_4;a,\mu)\,\frac{i}{\ell^2+m^2}\,.
\label{tadpole}
\end{eqnarray}
First we need
$\tilde\Gamma$ from the Feynman rule (\ref{Feynrule}) 
in accordance with the notations of Fig. \ref{fig:vertex}; that is 
we have to replace the incoming momenta $k_1$ and outgoing momenta $k_4$ 
by $k_3 \to -k_3=-\ell$ and $k_4 \to -k_4$. Thus, we have
\begin{eqnarray}
\label{Gamma}
&& \hspace{-1cm}
\tilde\Gamma(k_1,\ell,\ell,k^{out}_4;a,\mu) 
 = i(2\pi)^{n} \mu^{4-n}\;\frac{\lambda}{2}\, 
\Bigg\{1+ a_{\nu}\frac{1}{8}
\bigg[\bigg(k_{4\mu}k_{1\nu} - k_{1\mu}k_{4\nu} \bigg)\partial^{\ell}_{\mu}
\\
\nonumber
\hspace{-1cm}
&-&  
2\bigg({\ell}_{\mu}k_{4\nu} -3\ell_{\nu}k_{4\mu}
+8\delta_{\mu\nu}k_{4\rho}\ell_\rho\bigg)\partial^{k_1}_{\mu}
-2\bigg(\ell_{\mu}k_{1\nu}+k_{1\mu}\ell_{\nu}
-2\delta_{\mu\nu}k_{1\rho}\ell_\rho\bigg)\partial_\mu^{k_4}\\
\nonumber 
\hspace{-1cm}
&+&  
\bigg(\ell_{\mu}(2k_4-k_1)_{\nu}+\ell_{\nu}(2k_4-3k_1)_{\mu} - 
4 \delta_{\mu\nu} (k_4 - k_1)^{\rho}\ell_{\rho} 
 \bigg)\partial^{\ell}_{\mu} \bigg]
\Bigg\} \delta^{(n)}(k_1 - k_4)\,.
\nonumber
\end{eqnarray}
In order to obtain $\tilde\Gamma(k_1,\ell,\ell,k^{in}_4;a,\mu)$ we just have 
to replace $k_4 \to -k_4$ in (\ref{Gamma}).

As a next step, we compute $\Pi^{a}_2$ straightforwardly 
with the help of the notion that the 
integral (\ref{tadpole}) is an effective action describing 
the given one-loop quantum
process. So, employing integration 
by parts in (\ref{tadpole}) and
using dimensional regularization, we obtain
\begin{eqnarray}
\Pi^{0}_2&=&-\frac{\lambda}{2}\,I_0 \,,
\label{T0}\\
\Pi^{a\ne 0}_2&=&-\frac{\lambda}{2} \Bigg\{ \frac{3}{8} {(aK)}I_0 -
\frac{1}{4}{(aK)}\bigg( {\delta}_{\mu\nu} 
- \frac{a_{\mu}K_{\nu}}{(aK)}\bigg)I_{2,\mu\nu}\Bigg\}\,,
\label{Ta}
\end{eqnarray}
where $K=2k_4-k_1$. 
%The oscillator contribution from $\xi$-term 
%we obtain by crude approximation of (\ref{3}). That is, from
%\begin{eqnarray}
%\hspace{-.5cm}
%{\Pi}^{0,\xi\ne 0}_2=
%-i\int \frac{d^n\ell}{(2\pi)^n} \;
%\tilde\Gamma(k_1,\ell,\ell,k_4;a,\mu)\,
%\frac{\xi^2}{(\ell^2+m^2)^2}\partial_{\ell}^2 \delta^{(n)}(\ell)\,,
%\label{tadpolexi}
%\end{eqnarray}
%integrating by part, we found
%\begin{eqnarray}
%\Pi^{0,\xi\ne 0}_2&=&
%-\frac{\lambda}{2}\xi^2\frac{(8n)({\mu'}^2)^{4-\frac{n}{2}}}{m^6}
%\label{Txi}\,.
%\end{eqnarray}
In the above equations the presence of 
$(2\pi)^n \delta^{(n)}(k_1 - k_4)$ is understood, although not being explicitly stated.
For $n=4-\epsilon$, we have the well known integrals
\begin{eqnarray}
 I_0&=&(\mu^2)^{2-\frac{n}{2}} 
\int \frac{d^{{n}}\ell}{(2\pi)^n}\frac{1}{\ell^2 + m^2} 
=\frac{m^2}{(4\pi)^2} 
\Bigg[\Bigg(\frac{4{\pi}{\mu}^2}{m^2}\Bigg)^{\frac{\epsilon}{2}}
\Gamma(-1+\frac{\epsilon}{2})\;{{\Bigg]}_{\epsilon \to 0}} 
\nonumber\\ 
&=& \frac{-1}{8\pi^2} m^2   
\Bigg[\frac{1}{\epsilon}+\frac{\psi(2)}{2}+{\rm log}\sqrt{\frac{4\pi\mu^2}{m^2}} +...\Bigg]
\label{I0}\,,\\ 
I_{2,\mu\nu}&=& (\mu^2)^{2-\frac{n}{2}} 
\int \frac{d^n\ell}{(2\pi)^n}\frac{\ell_{\mu}\ell_{\nu}}{(\ell^2+m^2)^2}=
\frac{1}{2} \delta_{\mu\nu} I_0 \,,\;\;\; \delta_{\mu\nu}\delta_{\mu\nu}=n\,,
\label{I2}
\end{eqnarray}
with a simple pole at $\epsilon =0$.
Thus the expression (\ref{I0}) is divergent in the UV cut-off.

The non-vanishing contributions come from the commutative parts, 
that is from (\ref{T0}). 
%and from harmonic oscillator term
%(\ref{Txi}) which is finite for finite scalar field mass.
The integrals (\ref{I0}) and (\ref{I2}) for $n=4$ give $\Pi^{a\ne 0}_2=0$, producing 
%in the case $\xi=0$, 
the very well known commutative result (\ref{T0}). All contributions
proportional to $a$, coming from $\kappa$-Minkowski NC $\phi^4$ theory cancel out,
as one would naively expect by inspecting vertex (\ref{Gamma}). 

Clearly, the one loop computation has to be modified by
anticipating the momentum conservation on the $\kappa$-space.
To illustrate that something nonstandard appears in our model (\ref{Feynrule}), 
we start with the general one-loop integral (\ref{3}).
It should be noted that one cannot integrate over $k_3$ using 
the first delta $\delta^{(n)}(k_2 - k_3)$ -- from the propagator -- 
and replace $k_3$ by $k_2$ in the above
expression as it stands, because of the derivative with respect to $k_2$.
So, as a first step of computation we are using a simple trick: 
\begin{eqnarray}
\partial_\mu^{k_1} \delta^{(n)}(k_1 + k_2 - k_3 - k_4) &=& 
\partial_\mu^{k_2} \delta^{(n)}(k_1 + k_2 - k_3 - k_4)\,,
\nonumber\\
\partial_\mu^{k_1} \delta^{(n)}(k_1 + k_2 - k_3 - k_4) &=& 
-\partial_\mu^{k_3} \delta^{(n)}(k_1 + k_2 - k_3 - k_4)\,, \; {\rm etc.,}
\label{del3B}
\end{eqnarray}
and than we rewrite (\ref{Feynrule}) and (\ref{propagator}) as follows:
\begin{eqnarray}
 &&\tilde\Gamma(k_1,k_2,k_3,k_4;a)G(k_2,k_3) = 
\nonumber\\
&& 
\Bigg\{i(2\pi)^n\frac{\lambda}{4} 
\Bigg[1+a_\nu \bigg(2(-k_{1\mu}k_{2\nu} - k_{2\mu}k_{1\nu} + k_{3\mu}k_{4\nu} +
k_{4\mu} k_{3\nu} + 2 \delta_{\mu\nu} (k_{1\rho}k_{2\rho}-k_{3\rho}k_{4\rho}))
\nonumber\\
&&
+\frac{1}{2}(k_{1\mu}k_{3\nu}-k_{3\mu}k_{1\nu}+k_{2\mu}k_{3\nu}-k_{3\mu}k_{2\nu}
+ k_{1\mu}k_{4\nu} - k_{4\mu}k_{1\nu} + k_{2\mu}k_{4\nu}-k_{4\mu}k_{2\nu})\bigg)
\partial_\mu^{k_1}\Bigg] 
\nonumber\\
&&
\times\delta^{(n)} (k_1 + k_2 - k_3 - k_4)\Bigg\}\bigg[
\frac{i}{k_2^2+m^2} 
%\bigg(1-\frac{\xi^2}{(k_2^2+m^2)}\partial^2_{k_3}\bigg)
\delta^{(n)}(k_2-k_3)\bigg]
\,.
\label{Feynrule2}
\end{eqnarray} 
After performing the integration over $k_3$ in (\ref{Feynrule2}) 
we obtain the following expression:
\begin{eqnarray}
&&\Pi^{a \ne 0}_2= -\frac{\lambda }{8(2\pi)^n} ((ak_1)k_4-(ak_4)k_1)_{\mu}
\bigg[\partial_\mu^{k_1} \delta^{(n)}( k_1 - k_4 )\bigg]  
 \int \frac{d^n k_2}{k_2^2 + m^2}\,,
\label{del6B}
\end{eqnarray}
where $k_1$ and $k_4$ are the external momenta. 
This expression is quadratically divergent in the UV cut-off
representing the quantum loop modification of 
the free action (\ref{actioncomlex}), 
which can be nonzero because of the momentum conservation violation at the vertex. 
Due to the results \eqref{del6B}  
we obviously stumbled across the momentum nonconservation. 
Such results seem to favor our {\it hybrid} approach.

\subsubsection{Tadpole diagram: $\kappa$-deformed momentum conservation}

In the following computation of the 
tadpole diagram from Fig. \ref{fig:tadpol}, 
we fully implement the {\it hybrid} approach, 
that is the notion that standard momentum conservation is not satisfied.
%i.e. we use induced momentum conservation on $\kappa$-space represented 
%within delta functions in (\ref{delta}). 
However, in accordance with our {\it hybrid} approach,
at the end the undeformed momentum conservation law has to be applied. 
%General one-loop integral (\ref{3}) 
%can be roughly reduced to two terms, (\ref{tadpole}) and (\ref{tadpolexi}), respectively.  

In the next step, we are applying integration by parts. 
This of course plays an essential role in our {\it hybrid} approach.
Performing the computation
of all terms in the one-loop tadpole integral (\ref{tadpole})
with the Feynman rule (\ref{Feynrulekappa}) and the delta function (\ref{delta-}) and
for an arbitrary number of dimensions $n$, 
we obtain the following first order result:
\begin{eqnarray}
\Pi^a_2=\Pi^{0}_2 + \Pi^{a\ne 0}_2 &=& -\frac{\lambda}{2} 
\bigg[(1+(n-1)ak_1)+\frac{1}{2}(1-\frac{n}{4})(ak_1-2ak_4)\bigg]\,I_0 \,.
\label{0a0}
%\\
%\Pi^{0,\xi\ne 0}_2 + \Pi^{a\ne 0,\xi\ne 0}_2
%&=&-\frac{\lambda}{2}\xi^2\frac{8n({\mu'}^2)^{4-\frac{n}{2}}}{m^6}
%(1+(n-1)ak_1)\,.
%\label{a,x-i}
\end{eqnarray}
The first term in (\ref{0a0}) for $n =4$ corresponds to the result 
(\ref{T0}) from the previous subsection. 
From above formulas it is clear that there exist 
non-vanishing contributions even for $n=4$. They are arising 
from the $\kappa$-deformed momentum conservation rule, and entering through
the deformed $\delta$-function (\ref{delta-}) in the {\it hybrid} Feynman rule 
(\ref{Feynrulekappa}).
%, and from the harmonic oscillator 
%term in the action (\ref{actioncomlex})
%via the modified propagator (\ref{propagator1}).

For $n=4-\epsilon$, we obtain a modified expression 
for the tadpole in Fig. \ref{fig:tadpol}
in the limit $\epsilon \to 0$, where the $1/\epsilon$ divergence is explicitly isolated. 
For conserved external momentum in accordance with (\ref{delta-}), 
i.e. for $k_1=k_4\equiv k$, (\ref{I0}) and (\ref{0a0}) finally lead to, 
\begin{equation}
{\Pi}^{a}_2=\frac{\lambda m^2}{32\pi^2} \left[(1+3ak)\bigg(   
\frac{2}{\epsilon}+\psi(2)+{\rm log}\frac{4\pi\mu^2}{m^2}\bigg)
-\frac{9}{4}ak
% -(4-\epsilon)128\pi^2\frac{\xi^2 {\mu}^{4-\epsilon}}{m^8}
\right]\,.
\label{n=4}
\end{equation}
%where for simplicity we have used $\mu'=\mu$,
%and in $\xi$-term we retain the explicit $\epsilon$-dependence
%in order to keep dimensional and/or limiting procedure under control. 
The finite parts represent modifications of the scalar field self-energy
and depend explicitly on the regularization parameter, 
and the mass of the scalar field, 
%the magnitude of the translation invariance breaking, 
and it contains a correction $ak$ due to
the dependence on the energy $|k|$,
where the actual scalar field self-energy modifications occur.  
%Dependence of (\ref{n=4}) on $\kappa$-deformation parameter $a$
%enters explicitly, as we expected. 
%Note that there is no need to do renormalization at the point $(1+3ak) \to 0$.

The result (\ref{n=4}) is discussed next in the framework of Green's functions.
Generally we know that by summing all the 1PI contributions,
for the full free propagator (\ref{propagator}),
we get the following expression for the
two-point connected Green's function \cite{Casalbuoni}
\begin{eqnarray}
G_{(c,2)}^{a}(k_1,k_4)&=&%(2\pi)^n\delta^{(n)}(k_1-k_4)
\bigg[\frac{i}{k_1^2+m^2}+
\frac{i}{k_1^2+m^2}{\Pi}^{a}_2\frac{i}{k_1^2+m^2} +...\bigg]\,,
\label{Gc,2}
\end{eqnarray}
%where, symbolically, $m_1^2=m^2+\xi^2\partial^2_{k_1}\delta^{(n)}(k_1-k_4)$ 
%represents redefined mass.  
and as an illustration we resum the above series into
% the limit $\xi\to 0$:
\begin{eqnarray}
G_{(c,2)}^{a}(k_1,k_4)
 &{\longrightarrow}&\, 
 (2\pi)^n\delta^{(n)}(k_1 - k_4)\bigg[\frac{i}{k_1^2+m^2-{\Pi}^{a}_2}\bigg]\,.
\label{T2}
\end{eqnarray}

The genuine $1/\epsilon$ divergence in ${\Pi}^{a}_2$,
can only be removed by introducing the following counter term $\delta m^2$:
\begin{equation}
\delta m^2 \tilde \phi^*(k) \tilde \phi(k) = \frac{\lambda m^2}{32\pi^2}    
\Bigg[(1+3ak)\frac{2}{\epsilon} 
+f\bigg(\frac{4-\epsilon}{2},\frac{\mu^2}{m^2},ak\bigg)
\Bigg]\tilde \phi^*(k) \tilde \phi(k)\,,
\label{Rn=4}
\end{equation}
where $f$ is an arbitrary dimensionless function, fixed by renormalization conditions.
Adding the above counterterm contribution to the previous expression (\ref{Gc,2}) results 
%during the renormalization procedure  
in the shift $m^2 \to m^2 + \delta m^2$  in (\ref{Gc,2})/(\ref{T2}), thus leading to
\begin{eqnarray}
{\tilde G}_{(c,2)}^{a}(k_1,k_4)&=&
\Bigg[G_{(c,2)}^{a}(k_1,k_4)+
G_{(c,2)}^{a}(k_1,k_4)(-\delta m^2)G_{(c,2)}^{a}(k_1,k_4) +...\Bigg]
%_{\xi\to 0}
\nonumber\\
 &{\longrightarrow}& (2\pi)^n\delta^{(n)}(k_1-k_4)
\bigg[\frac{i}{k_1^2+m^2+\delta m^2  
 -{\Pi}^{a}_2}\bigg]\,,
\label{T2m}
\end{eqnarray} 
where ${\tilde G}_{(c,2)}^{a}$ denotes the Green's function 
including the contribution from the counter term. % incorporated.

However, since ${\Pi}^{a}_2$ was computed for the
free propagator (\ref{propagator}), it is consistent to
compute the two-point connected Green's function
under the same approximation.
After the resummation of (\ref{Gc,2}), using of (\ref{propagator}), we find
\begin{eqnarray}
G_{(c,2)}^{a}(k_1,k_4)
=%(2\pi)^n\delta^{(n)}(k_1 - k_4)
\frac{G}{1-G\;{\Pi}^{a}_2} \,.
\label{Gexp}
\end{eqnarray}
In order to identify the proper counter term for the above expression, we resum the 
series in (\ref{T2m}) with the full free propagator $i/(k_1^2+m^2)$
and replace (\ref{Gc,2}) $\to$ (\ref{Gexp}):
\begin{eqnarray}
{\tilde G}_{(c,2)}^{a}(k_1,k_4)
=%(2\pi)^n\delta^{(n)}(k_1 - k_4)
\frac{G}{1+G(\delta m^2-{\Pi}^{a}_2)} \,.
\label{tilGexp}
\end{eqnarray}
In (\ref{tilGexp}), $\delta m^2$ is a generic quantity. 
This is due to the fact that expression (\ref{n=4}) contains the finite parts too.
The requirement $\delta m^2={\Pi}^{a}_2$ removes the infinity.
Thus, we have
\begin{eqnarray}
{\tilde G}_{(c,2)}^{a}(k_1,k_4)
=\frac{G}{1+G\frac{\lambda m^2}{32\pi^2}
\bigg[f-(1+3ak)\bigg(\psi(2)+{\rm log}\frac{4\pi\mu^2}{m^2}
-\frac{9}{4}\frac{ak}{1+3ak} \bigg)\bigg]} 
\,.
\nonumber\\
\label{tilGexpf}
\end{eqnarray}

Precise extraction and removal of the genuine UV divergence is performed next via 
(\ref{Rn=4}) in the context of the discussion of 
${\tilde G}_{(c,2)}^{a}(k_1,k_4)$ for different energy regimes; 
that is from low energy to extremely high -Planck scale- energy propagation.

There is a very interesting property of expression (\ref{n=4}) at extreme energies. 
Namely, there exists a term $(1+3ak)$ which for $(1+3ak \to 0)$ tends to zero linearly.
For low energies and/or small $\kappa$-deformation $a$, i.e. 
for $ak \simeq 0$, (equivalent to $a\simeq 0$), which is far away from the point $(1+3ak=0)$, 
this is not the case. \\

\noindent
{\it Low energy limit}\\
Using the finite combination $(\delta m^2 - \P^{0}_2)$ 
for low energy ($ak = 0$), and at the order $\lambda$ 
\begin{equation}
\delta m^2-{\Pi}^{0}_2=\frac{\lambda m^2}{32\pi^2}    
\Bigg[f-\psi(2)-{\rm log}\frac{4\pi\mu^2}{m^2}\Bigg]\,,
\label{Rn4}
\end{equation}
we get from (\ref{n=4}) and (\ref{tilGexpf}): 
%in addition to removing spurious term
%$\xi^2\partial^2_{k}\delta^{(4)}(k)$, 
\begin{eqnarray}
{\tilde G}_{(c,2)}^{0}(k_1,k_4)
=\frac{G}{1+G\frac{\lambda m^2}{32\pi^2}
\bigg(f-\psi(2)-{\rm log}\frac{4\pi\mu^2}{m^2}
% +512\pi^2\frac{\xi^2 {\mu}^{4}}{m^8}
\bigg) } \,.
\nonumber\\
\label{tildeGc2}
\end{eqnarray}
%while in the case $\xi=0$
%\begin{eqnarray}
%{\tilde G}_{(c,2)}^{0,0}(k_1,k_4)
%=\frac{(2\pi)^4\delta^{(4)}(k_1-k_4)}{k_1^2+m^2\bigg(1+\frac{\lambda}{32\pi^2}
%\bigg[f-\psi(2)-{\rm log}\frac{4\pi\mu^2}{m^2}\bigg]\bigg)} \,.
%\nonumber\\
%\label{tildeGc200}
%\end{eqnarray}
This expression has a pole in Minkowski space, and we can define 
the renormalization condition by requiring
that the inverse propagator at the physical mass isequal to $k_1^2 + m^2_{phys/low}$.
This choice 
determines uniquely the sum of the residual terms in (\ref{Rn4}),
which is in accordance with the commutative $\phi^4$ theory result \cite{Casalbuoni}.\\

\noindent
{\it Planckian energy limit}\\
At the limiting point $(1+3ak\to 0)$, which corresponds 
to the extreme energies $|k|$,
where the components of the $\kappa$-deformation parameter $a_\mu$ are extremely small,
of order Planck length, the divergence in (\ref{n=4}) is removed
{\it under the choice that $(1+3ak)$ tends to zero linearly, i.e. with 
the same speed as $\epsilon$ does}. 
That is, in the Planckian energy limit 
\begin{equation}
\frac{(1+3ak)\to 0}{\epsilon \to 0} \longrightarrow {\cal O}(1)\,,
\label{aklimit}
\end{equation}
the $1/\epsilon$ and $ak$ terms, from (\ref{n=4}), do contribute.
  
Assuming that our $\kappa$-noncommutativity is spatial
$a_{\mu}=(\vec a,0)$, and using the momentum
along the third axis $k_{\mu}=(0,0,E,iE)$,
i.e. for $ak=Ea_3$, Eq. (\ref{n=4}) in the Planckian energy 
limit (\ref{aklimit}) gives 
\begin{equation}
\Pi^{a}_2\,\bigg|_{(3E a_3 +1\to 0)} 
\longrightarrow\,\frac{\lambda m^2}{32\pi^2}\;
\bigg[2-\frac{9}{4}ak\bigg]_{(3E a_3 +1\to 0)}\,,
\label{ak=1/3}
\end{equation}
producing the following modified Green's function (\ref{T2}):
\begin{equation}
\Pi^{a}_2\,\bigg|_{(3E a_3 +1\to 0)}\to 
\frac{\lambda m^2}{32\pi^2}\,\frac{11}{4}\,
\Rightarrow\;
{\tilde G}_{(c,2)}^{a}(k_1,k_4)\bigg|_{(3E a_3 +1\to 0)}\simeq 
\frac{i(2\pi)^4\delta^{(4)}(k_1-k_4)}
{k_1^2+m^2\bigg(1-\frac{\lambda}{32\pi^2}\frac{11}{4}\bigg)}\,.
\label{11/4}
\end{equation}
At the exact zero-point $3E a_3+1=0$ however,  
we obtain a different result from (\ref{n=4}):
\begin{equation}
\Pi^{a}_2\,\bigg|_{(3E a_3 +1= 0)}= 
\frac{\lambda m^2}{32\pi^2}\,\frac{3}{4}\,\;\;
\Rightarrow\;\;
{\tilde G}_{(c,2)}^{a}(k_1,k_4)\bigg|_{(3E a_3 +1= 0)}\simeq 
\frac{i(2\pi)^4\delta^{(4)}(k_1-k_4)}
{k_1^2+m^2\bigg(1-\frac{\lambda}{32\pi^2}\frac{3}{4}\bigg)}\,.
\label{3/4}
\end{equation}

For the time-space noncommutativity with the time component of 
the $\kappa$-deformation parameter $ia_0$ being also
of the Planck length order and vanishing space component, 
i.e. for $ak=-Ea_0$, Eq. (\ref{n=4}) in the Planckian energy 
limit (\ref{aklimit}) produces the result
\begin{equation}
\Pi^{a}_2\,\bigg|_{(3E a_0 -1\to 0)} 
\longrightarrow\,\frac{\lambda m^2}{32\pi^2}\;
\bigg[2-\frac{9}{4}ak\bigg]_{(3E a_0 -1\to 0)}\,,
\label{3/4/}
\end{equation} 
equivalent to (\ref{ak=1/3}), 
leading to the final results identical with (\ref{11/4}) and (\ref{3/4}).

The existence of the linear type of the limit $(1+3ak \to 0)$ which removes the genuine UV
divergence $1/\epsilon$ is a new, previously unknown feature of
NC $\kappa$-Minkowski $\phi^4$ theory at linear order in $a$.
The above expressions give the $\kappa$-deformed dispersion relations.
Mass shift receives the fixed value (\ref{ak=1/3}),  
independent of the function $f$, but it does depend on the scalar factor $ak$;
that is, it depends on the direction, 
on the energy $|k|$ and on the $\kappa$-deformation parameter $a$. 
Thus, equations (\ref{ak=1/3})/(\ref{3/4/}) and (\ref{11/4})/(\ref{3/4}) represent 
a birefringence \cite{Abel:2006wj,Buric:2010wd} of the massive scalar field mode. 
Namely, both of the finite terms in (\ref{tilGexpf})
do not contribute in the limit $(1+3ak) \to 0$, and the possibility of
their internal cancellation is diminished. 
%This way massive scalar field mode birefringence arises as genuine effect. 
The inverse propagator determines the physical mass $m^2_{phys/Planck}$
at Planck scale energies.

\section{Discussion and conclusion}

A description and discussion of the paper's main results are in order.\\
%First, we point out  the most important new results:\\
(i) The integral measure problems on $\kappa$-Minkowski spacetime 
are avoided by the introduction of  
the new $\kappa$-deformed ${\star}_h$-product (\ref{newstarproduct}),
which naturally absorbes the measure function due 
to the hermitian realization (\ref{2.10d}).  \\
\noindent
(ii) The trace/integral identity 
(\ref{property}) is valid for $\kappa$-deformed spaces, but only if one
is dealing with the hermitian realization (\ref{2.10d}), as one should,
because only hermitian realizations have physical meaning.
Due to the integral identity (\ref{property}), the only deformation with respect to the
standard scalar field action comes from the interaction term in (\ref{actionnew}).\\
\noindent
(iii) %The action (\ref{actioncomlex}) includes an 
%harmonic type of the interaction term and is expanded up to first order in the 
%deformation parameter $a$, 
%producing an effective theory on commutative spacetime.
%Despite $\kappa$-deformation mixture with $\xi^2$ term in (\ref{actionnew})
%via ${\star}_h$ product, at first order in $\kappa$-deformation these two
%features of our model separate completely in (\ref{actioncomlex}).
The action (\ref{actioncomlex}) produces modified 
equations of motion (\ref{eom2}) 
and conserved deformed currents (\ref{current})
due to the internal symmetry satisfied at that order.
%The above properties are very welcome.
\\
\noindent
(iv) The truncated $\kappa$-deformed action (\ref{actioncomlex})
does not possess the celebrated UV/IR mixing \cite{Grosse:2005iz}.
The lack of the UV/IR mixing is a general feature of most of 
NCQFT expanded in terms of the noncommutative deformation parameter.
%However, resummation of expanded action could in principle restore
%nonperturbative character of the model 
%(see for example \cite{Horvat:2011iv}), thus restoring
%presence of UV/IR mixing in quantum loop computations.
%Under such circumstances UV/IR mixing would help
%in determining what the UV theory might be, that is, it would help to
%determine UV completeness of the theory.
%The UV/IR mixing connects NCGFT with Holography, via UV and IR cutoffs 
%in a model independent way \cite{Horvat:2010km}, thus representing windows to quantum
%gravity phenomena \cite{Horvat:2010km,Cohen:1998zx,Szabo:2009tn}.
\\
\noindent
(v) Next, we discuss the result for the tadpole diagram  
contribution to the propagation
and/or self-energy of our scalar field $\phi$ 
for arbitrary number of dimensions $n$, depicted in Fig. \ref{fig:tadpol}, 
as a function of the $\kappa$-deformed momentum conservation law. This originates from
deformed statistics on the $\kappa$-Minkowski spacetime.
Our approach is a kind of {\it hybrid} approach modeling between standard QFT and NCQFT on 
the $\kappa$-Minkowski spacetime, involving $\kappa$-deformed $\delta$-functions
in the Feynman rules.\\
\noindent
(vi) When we worked with the standard conservation of momenta
and the undeformed $\delta$-function, the contributions to 
the tadpole diagram to first order in $a$ are zero. 
The deformed $\delta$-function can be written in terms 
of a leading term plus corrections in $a$. 
Since the Feynman rule (\ref{Feynrule}) has already terms linear in $a$, 
we have to retain only the zeroth order term in
the modified $\delta$-function, because otherwise we get 
terms of quadratic and higher orders in $a$ 
(and this is not what we are interested in). 
The only term where we need to take into account 
the $\delta$-function correction linear in $a$, is the leading order term 
in the Feynman rule (\ref{Feynrule}). However, this term 
is vanishing due to integration over the loop momentum.
%Analyzing (\ref{del6B}), we recognize that 
%something nonstandard appears in the model 
%for the tadpole integral at first order in $\lambda$ and $\kappa$-deformation $a$.
\\
\noindent
(vii) It appears that in the computation of the tadpole diagram integrals  
for arbitrary number of dimensions $n$, 
for $n=4$ all contributions linear in $a$ canceled each other automatically. 
For dimensions $n\not= 4$, the same linear in $a$ contributions become nonzero,
regardless which momentum conservation rule is used. 
%However the harmonic oscillator term from the action (\ref{actioncomlex})
%modifies the mass term in the free propagator (\ref{propagator}), thus 
%producing the additional contribution (\ref{Txi}) from the tadpole in Fig. \ref{fig:tadpol}.
The propagation of the scalar field $\phi$ for  
$n=4$ dimensions also receives a modification from
the $\kappa$-deformation at linear order in the deformation parameter $a$. 
%and it does receive
%contribution due to oscillator term in the action for any number of dimensions.
\\
\noindent
(viii) In the final computation of 
the tadpole diagram depicted in Fig. \ref{fig:tadpol}, we fully implement the notion
of our {\it hybrid} approach,
%that is that standard momentum conservation is not explicitly satisfied
i.e. we have to use the momentum conservation on $\kappa$-space given in (\ref{ominus}), 
while at the end of the computation undeformed momentum conservation has to be applied. 
We have found non-vanishing contributions even in $n=4$ dimensions.
They are arising from the
%from the harmonic oscillator term in the action (\ref{actioncomlex})
%via modified propagator (\ref{propagator1}), and 
$\kappa$-deformed momentum conservation rule entering through
the deformed $\delta$-function in the {\it hybrid} Feynman rule (\ref{Feynrulekappa}).
We have found a fully modified expression for the tadpole in Fig. \ref{fig:tadpol}
in the limit $\epsilon \to 0$, where the genuine $1/\epsilon$ (UV) 
divergence is explicitly isolated. 
For conserved external momenta, i.e. for $k_1=k_4\equiv k$, we obtain 
the two-point function (\ref{n=4}),
where the finite parts represent the modification 
of the scalar field self-energy $\Pi^{a}_2$
and depend explicitly on the regularization parameter $\mu^2$ and
the mass of the scalar field $m^2$.
%, and the magnitude of translation invariance breaking $\xi^2$
The most important is that 
(\ref{n=4}) contains the finite correction $ak$
due to the deformed statistics on the $\kappa$-Minkowski spacetime,
thus, we obtain an explicit dependence on 
the direction of the propagating energy, its scale $|k|=E$ 
and the $\kappa$-deformation parameter $a$, as we expected.\\
\noindent
(ix) The two-point function (\ref{n=4}) is next applied in the framework 
of the two-point connected Green's function
for three energy regimes, that is for low energies, for Planck scale energies,
and for intermediate energies, respectively.
For low energy scale and/or small $\kappa$-deformation $a$, i.e. 
for $ak\simeq 0$, which is far away from the point $(1+3ak=0)$, 
the $\kappa$-deformation dependence of the 
two-point Green's function completely drops out (\ref{tildeGc2}). 
The genuine UV divergence in (\ref{n=4}) 
has been removed by subtracting the counterterm $\delta m^2$ (\ref{Rn=4}),
from previous contribution (\ref{Gc,2}),
or through shifting $m^2$ into $(m^2 + \delta m^2)$ in (\ref{T2}).
In this case the mass shift (\ref{tildeGc2}) could increase or decrease $m^2$
depending on the values of the function $f$. \\ 
\noindent
(x) For energies within the limits $\frac{-1}{3a_3} \ll E \ll 0$ (or equivalently
$\frac{1}{3a_0} \gg E \gg 0$), the full expression (\ref{n=4}), 
with the mass counterterm (\ref{Rn=4}) has to be used
in order to determine the Green's function (\ref{T2m}). 
%In that particular case
%the harmonic oscillator term in (\ref{n=4}) will also 
%give a non-negligible contribution. It's coupling $\xi^2$
%could be in principle determined via higher order contributions 
%to the Green's function, which is certainly an issue to be addressed
%in future work.  
\\
\noindent
(xi) At the Planckian energy scale, due to the existence of 
linear type of limits $(1+3ak) \to 0$, we have a new situation and
distinguish two cases. They both are new, previously unknown features of
the linear order in $a$ NC $\kappa$-Minkowski $\phi^4$ model.
In the first case we have the limit (\ref{aklimit}) which produces
the self-energy and/or the modified Green's function (\ref{11/4}).\\
\noindent
(xii) Second case, that is the exact zero-point where $1+3E a_3=0$, 
represents in fact a genuine type of the zero-point which  
exactly removes the UV divergence $1/\epsilon$,
producing the self-energy and/or modified Green's function (\ref{3/4}).
In both cases the mass term is shifted in the same direction 
({\it the same sign!}) but by a different amount, $+11/4$ versus $+3/4$, respectively.
Or more precisely, we can say that the mass shift during the limiting process  
$(3E a_3 +1\to 0)$ drops from the value proportional to 
$+11/4$ to the exact value proportional to $+3/4$.\\ 
\noindent
(xiii) The results (\ref{11/4}) and (\ref{3/4}) are the same for 
two different choices of $\kappa$-noncommutativity, 
%(with appropriate choice of referent
%system for momentum $k_{\mu}$), 
i.e. for choice $a_{\mu}=(0,0,a_3,0)$, or $a_{\mu}=(0,0,0,ia_0)$, respectively.\\
\noindent
(xiv) At the Planckian propagation energy scale $E \simeq \frac{-1}{3a_3}$,  
the contribution of the tadpole in Fig. \ref{fig:tadpol} tends to 
the fixed finite value, between (\ref{11/4}) and (\ref{3/4}), respectively. 
Due to effects of the $\kappa$-Minkowski statistics, this only
depends on the direction of the propagation and the 
$\kappa$-deformation parameter $a$.
In this way (\ref{ak=1/3})/(\ref{3/4/}) and (\ref{11/4})/(\ref{3/4})   
represent $\kappa$-deformed dispersion relations,
producing a genuine birefringence, \cite{Abel:2006wj,Buric:2010wd}, of 
the massive scalar field modes, 
%which arises as genuine effect at the first order in 
%$\kappa$-deformation parameter $a$. 
similarly to the chiral fermion field
birefringence in the truncated model \cite{Buric:2010wd}.\\  
\noindent
(xv) Considering full renormalization, besides the $\delta m^2$ counterterm, 
also the other divergent parts have to be added as
counterterms to the free Lagrangian (\ref{actioncomlex}):\\ 
$ 
\int\,({\cal L} + {\cal L}_{ct}) = S[\phi_B, m_B, \lambda_B, a_B]\,,
$
where the index $B$ denotes bare quantities.
That would include an analisys of the $4-point$ one-loop contributions, 
counterterms $(\mu^2)^{2-\frac{n}{2}}\delta\lambda$ 
%and $({\mu}^2)^{4-\frac{n}{2}}\delta\xi$,
as well as 2-loop expansion for the 2-point Green's function with
insertion of the counterterms in multi-loop diagrams.
Certainly, the full analysis of the renormalization group equations 
should be studied following the same lines.
However, the full renormalization of our action (\ref{actioncomlex}) 
is anyhow beyond the scope of this paper and it is planned for our next project. 

%\noindent
Regarding the effects of the statistics according to the described arguments, 
we repeat that within first order 
in the $\kappa$-deformation $a$, {\it the statistics
effects on the $\kappa$-Minkowski in our hybrid model do arise as
semiclassical/hybrid behavior of the first order quantum correction,
thus showing birefringence of the massive scalar field mode.}
%We believe that this property of such a constructed model, is of 
%importance for further possible research towards quantum gravity. 
%At higher orders in $a$ the
%matter would become growingly interesting and complicated.

\noindent
%{\bf Acknowledgment}\\
\acknowledgments
We would like to acknowledge A. Andra\v si, A. Borowiec, H. Grosse, J. Lukierski, 
V. Radovanovi\' c and J. You for fruitful discussions.
We would specially like to thank to J. Lukierski and A. Borowiec
for careful reading of the manuscript and numerous valuable remarks which we 
incorporated into the final version of this manuscript.
J.T. would like to acknowledge support of Alexander von Humboldt Foundation
(KRO 1028995), and Max-Planck-Institute for Physics,
and W. Hollik for hospitality. %We thank G. Duplan\v ci\' c for drawing graphs.
This work was supported by the Ministry of Science and Technology of
the Republic of Croatia under contract No. 098-0000000-2865 and 098-0982930-2900.
% The work of J.T. is supported in part by EU (HEPTOOLS) 
% project under contract No. MRTN-CT-2006-035505. 

%%%%%%%%%%%%%%%%%%%%%%%%%%%%%%% bibliography%%%%%%%%%%%%%%%%%%%%%%%%%%%%%%%
 %%%%%%%%%%%%%%%%%%%%%%%%%%%%%%%%%%%%%%%%%%%%%%%%%%%%%%%%%%%%%%%%%%%%%%%%%

\end{document}